\documentclass[12pt]{article}
 \usepackage{scicite}
 \usepackage{float}
\usepackage{placeins}
 \usepackage{url}
 \usepackage{lscape}
 
\def\aap{Astron. \& Astrophys.}                
\newcommand{\aj}{Astron. J.}
\newcommand{\apjs}{Astrophys. J.}
\newcommand{\apj}{Astrophys. J.}
\newcommand{\apjl}{Astrophys. J.}
\newcommand{\mnras}{Mon. Not. R. Astro. Soc.}
\newcommand{\pasa}{Publ. Astron. Soc. Australia}
\newcommand{\nat}{Nature}

\newcommand{\procspie}{Proc. SPIE.}
\newcommand{\pasp}{Pub. Astron. Soc. Pacific}
\newcommand{\annrevastro}{Annu. Rev. Astron. Astrophys.}
\newcommand{\pccm}{\ensuremath{\mathrm{pc~cm}^{-3}}}
\newcommand{\stellarmass}{2.2 \times 10^{10} {\rm M}_\odot} 
\newcommand{\reffv}{2.8 \pm 0.2} 
\newcommand{\dmcosmic}{307} 
\newcommand{\dmexcess}{46} 

\usepackage{graphicx}
\usepackage{amsmath,amssymb} 
\usepackage{times}

\newenvironment{sciabstract}{%
\begin{quote} \bf}
{\end{quote}}

\begin{document}

\vspace{0.2in}

\noindent
{\bf Title:} A single fast radio burst  localized to a  massive galaxy at cosmological distance \\


\noindent
{\bf Authors:}
K.~W. Bannister$^{1,\ast}$, 
A.~T.~Deller$^{2}$, 
C.~Phillips$^{1}$, 
J.-P.~Macquart$^{3}$,
 J.~X.~Prochaska$^{4,5}$,
 N.~Tejos$^6$, 
  S.~D.~Ryder$^7$,
  E.~M.~Sadler$^{1,8}$,
  R.~M.~Shannon$^{2,\ast}$,
   S.~Simha$^4$, 
   C.~K.~Day$^2$,
    M.~McQuinn$^9$,
     F.~O.~North-Hickey$^3$
      S.~Bhandari$^1$
W.~R. Arcus$^{3}$, 
V.~N.~Bennert$^{10}$,
 J.~Burchett$^6$,
M.~Bouwhuis$^{1,11}$, 
R.~Dodson$^{12}$, 
R.~D.~Ekers$^{1,3}$, 
W.~Farah$^{2}$, 
C.~Flynn$^{2}$, 
C.~W. James$^{3}$, 
M.~Kerr$^{13}$, 
E.~Lenc$^{1}$, 
E.~K.~Mahony$^{1}$, 
J.~O'Meara$^{14}$, 
S.~Os{\l}owski$^{2}$,
H.~Qiu$^{1,8}$, 
T.~Treu$^{15}$, 
V.~U$^{16}$, 
T.~J. Bateman$^{8}$, 
D.~C.-J.~Bock$^{1}$, 
R.~J.~Bolton$^{1}$, 
A.~Brown$^{1}$, 
J.~D.~Bunton$^{1}$,
A.~P. Chippendale$^{1}$, 
F.~R. Cooray$^{1}$, 
T.~Cornwell$^{17}$,
N.~Gupta$^{18}$, 
D.~B.~Hayman$^{1}$, 
M.~Kesteven$^{1}$, 
B.~S.~Koribalski$^{1}$, 
A.~MacLeod$^{1}$, 
N.~M.~McClure-Griffiths$^{19}$, 
S.~Neuhold$^{1}$,
R.~P.~Norris$^{1,20}$,
M.~A.~Pilawa$^{1}$,
R.-Y.~Qiao$^{1}$, 
J.~Reynolds$^{1}$,
D.~N. Roxby$^{1}$, 
T.~W. Shimwell$^{21}$,
M.~A.~Voronkov$^{1}$, 
C.~D. Wilson$^{1}$

\noindent
\footnotesize{$^\ast$ To whom correspondence should be addressed. E-mail:  keith.bannister@csiro.au (K.W.B.), rshannon@swin.edu.au (R.M.S.). Affiliations are listed at the end of the main text. }

\begin{center}
{\normalsize
Accepted: 19 June 2019 }
\end{center}





\baselineskip 24pt

\begin{sciabstract}
Fast Radio Bursts (FRBs) are brief radio emissions from distant astronomical sources. Some are known to repeat, but most are single bursts. Non-repeating FRB observations have had insufficient positional accuracy to localize them to an individual host galaxy. We report the interferometric localization of the single pulse FRB~180924 to a position 4~kpc from the center of a luminous galaxy at redshift $0.3214$. The burst has not been observed to repeat. The properties of the burst and its host are markedly different from the only other accurately localized FRB source.  The integrated electron column density along the line of sight closely matches models of the intergalactic medium, indicating that some FRBs are clean probes of the baryonic component of the cosmic web.
\end{sciabstract}

Cosmological observations have shown that baryons comprise $4\%$ of the energy density of the Universe, of which only about 10\% is in cold gas and stars \cite{Fukugita04}, with the remainder residing in a diffuse plasma surrounding and in between galaxies and galaxy clusters. 
The  location and density of this material has been challenging to characterize, and up to 50\% of it remains unaccounted  \cite{Shull12}. 

Fast radio bursts (FRBs; ref. \cite{Thornton13}) are bright bursts of radio waves with millisecond duration. They can potentially be used to detect, study, and map this medium,
as bursts of emission are dispersed and scattered  by their passage through any ionized material, including the intergalactic medium. If the emission is linearly polarized and any of the media are magnetized, the burst is also subject to Faraday rotation, i.e., the frequency dependent rotation of the plane of linear polarization due to its passage through a magnetized plasma \cite{Masui15}.

Detailed studies of the medium, and the bursts themselves, require localization of bursts to host galaxies, so that burst redshifts and their propagation distances can be determined.

To date, only one source (FRB~121102) has been localized \cite{Chatterjee17}  to sufficient accuracy to identify a host. Is is also one of only two FRBs known to repeat \cite{Chimerepeater}.
The burst localization  was made through radio-interferometric detections of repeated bursts.
The burst source lies in a luminous radio nebula \cite{Chatterjee17} within a dwarf galaxy with high star formation rate per unit stellar mass, at redshift $z=0.19$ \cite{Tendulkar17}. 
This has led to the hypothesis that bursts are produced by young magnetars embedded in pulsar wind nebulae \cite{Margalit18}, with the host galaxy properties suggesting an indirect connection between FRBs and other transient events which are common in this type of galaxy, such as superluminous supernovae and long-duration gamma-ray bursts.  

The relationship between the source of FRB~121102 and the larger FRB population is unclear\cite{Nicholl17,Shannon18,Ravi19}.   
 Many sources have not been observed to repeat despite extensive campaigns spanning hundreds to thousands of hours \cite{Ravi16,Shannon18}.
The progenitors and mechanism by which burst emission is generated remain uncertain.
Localizing examples of further bursts, including those from a population that have not repeated, is required to determine their nature and establish if they can be used as cosmological probes.

\section*{Localizing fast radio bursts with ASKAP} 

The Australian Square Kilometre Array Pathfinder (ASKAP, Ref. \cite{Johnston08}), a $36$-antenna radio interferometer, has a specially designed mode capable of directly localizing dispersed pulses, such as fast radio bursts \cite{Clarke14}.
Each of the $12$-m antennas has been placed in a quasi-random configuration with baselines extending to $6$-km lengths, resulting in a maximum angular resolution of $10$~arcsec at a frequency of $1320$~MHz, enabling positions to be measured to a statistical precision of $\sim$$10$~arcsec/(2$\times$S/N),  where S/N is the source signal to noise ratio. 

The antennas are equipped with phased-array feed receivers \cite{HayOSullivan2008}, each of which can form $36$ simultaneous dual-polarization beams on the sky using digital beamforming, producing a total field-of-view of $\sim 30$~deg$^2$.
For burst detection, the beamformers produces channelized autocorrelation spectra for both linear polarizations of all beams, with an integration time of $864$\,$\mu$s and channel bandwidth of $1$~MHz in these observations.
We used $336$ channels centered at $1320$\,MHz.
A real-time detection pipeline incoherently adds the spectra from all available antennas ($24$ antennas in these observations) and polarization channels, then searches \cite{Bannister17} the result for dispersed pulses \cite{online}.

Burst localization is completed with a second data product that utilizes both the amplitude and phase information of the burst radiation.   
The beamformers store samples of the complex electric field for all beams and both polarizations in a ring buffer of $3.1$~s duration, with the oldest data being continuously overwritten by new data. 
 The data are saved for offline interferometric analysis only when the pipeline identifies a candidate.  
 For the searches reported here the triggering required pulses with  widths less than $9$\,ms and ${\rm S/N} > 10$. 

Previous searches with ASKAP used antennas pointed in different directions to maximize sky coverage \cite{Bannister17,Shannon18}. In contrast, our observations used antennas all pointed in the same direction, enabling the array to act as an interferometer capable of sub-arcsecond localization with a $30$~deg$^{2}$ field of view.
We targeted high Galactic latitude fields (Galactic latitude $|b| \sim 50^\circ$), that had been observed previously \cite{Bannister17,Shannon18}, and Southern circumpolar fields. 
The high-latitude fields were observed regularly through 2017 and early 2018  for a total duration of $>12,000$~hr \cite{Shannon18}   enabling us to put constraints on burst repetition.
For daytime observations, circumpolar fields were observed to enable prompt follow-up from Southern-hemisphere optical telescopes.  

\section*{The detection of FRB~180924} 

\begin{figure}
    \centering
    \begin{tabular}{c}
   \includegraphics[scale=0.65]{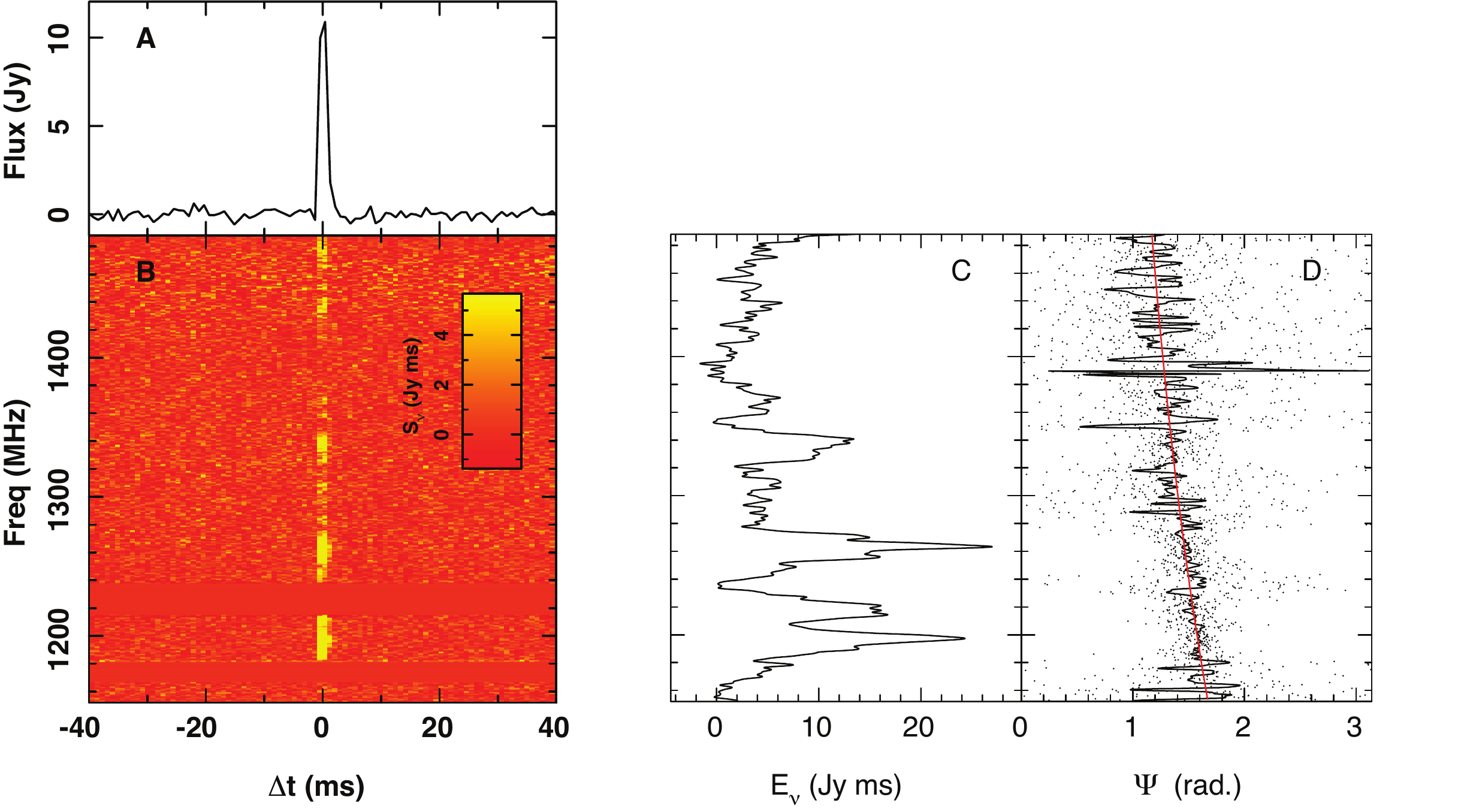}  
    \end{tabular}
    \caption{{\bf Spectral and polarimetric properties of FRB~180924.} (A) Integrated pulse profile. (B) Burst discovery dynamic spectrum, dedispersed by the measured dispersion measure (DM=$361.42$ \pccm).    The white bands are regions flagged due to radio-frequency interference in the high time resolution data. (C) Burst fluence spectrum (E$_\nu$) averaged over the pulse.  For this lower time resolution spectrum we partially mitigated the radio interference, so present estimates of the spectrum in the affected part of the bands flagged in (A) \cite{online}.  (D)  Polarization position angle ($\Psi$) of the burst.  The dots are measurements for individual spectral channels.  The black curve shows a version smoothed using a Gaussian kernel with a standard deviation of $5$ channels. The red line is the maximum-likelihood model for the polarization position angle swing of the burst assuming RM$=14$~rad\,m$^{-2}$. \cite{online}.}
    \label{fig:FRB180924}
\end{figure}

We detected a burst (FRB~180924, see Fig. \ref{fig:FRB180924}), with a signal to noise ratio  of $21$ in one of the high Galactic latitude fields. 
The search pipeline identified the burst $281$~ms after the dispersed pulse swept across the lowest frequency channel, and triggered the download of the buffer containing the burst. 

The properties of the burst, listed in Table \ref{tab:burst_properties}, and the strong spectral modulation (see Fig. \ref{fig:FRB180924}B), are  similar to the previous examples detected with ASKAP in lower-sensitivity searches \cite{Shannon18,Macquart18}, suggesting that they belong to the same population.   
The dispersion measure (DM) of the burst, which is the integrated free electron content along the line of sight,  weighted by the rest frame frequency while passing through the dispersing medium, is  $361.42 \pm 0.06$\,pc\,cm$^{-3}$ and the burst fluence is $16 \pm 1$~Jy\,ms (1~Jy = $10^{-26}$ W~Hz$^{-1}$~m$^{-2}$).  
The burst is $80\%$ linearly polarized and shows evidence for only modest Faraday rotation.
The measured strength of the Faraday rotation (the rotation measure - RM) is  RM=$14\pm1$\,rad\,m$^{-2}$ \cite{online}.
The Galactic foreground contribution to the Faraday rotation along high latitude lines of sight is low; the Milky Way Faraday rotation along this line of sight is predicted\cite{Oppermann12} to be $7.5~\mathrm{rad~m^{-2}}$.
The pulse shows evidence for scatter broadening with a scattering time scale $\tau_s=580 \pm 20~\mu$s at a frequency of $1.2$~GHz \cite{online}.  
 
\section*{A sub-arcsecond localization}

We localized the burst  using an image made from the $3.1$~s of voltage data, produced using  techniques developed for long-baseline radio interferometry.
Two teams blindly analyzed the data using different pipelines and codes, and derived the same initial source positions \cite{online}.
In a refined, coherently formed, optimally weighted image,  the burst was detected with a signal to noise ratio of $194$, from which the position was measured with a statistical uncertainty (from thermal noise alone) of $0.04$~arcsec.

To identify a host galaxy it is necessary to tie the radio image to an optical reference frame.   
We register the position of the burst on a deep Dark Energy Survey (DES, Ref. \cite{DESDR1}) optical image of the region 
by bootstrapping the radio-interferometric image of the burst to a deeper radio observation of the field that can subsequently be referenced to a standard sky coordinate system  \cite{online}. 
In addition to the burst, three constant (non-transient) radio sources were also detected in our $3.1$-s ASKAP image. 
We compared their measured positions with those obtained from phase-referenced observations with the Australia Telescope Compact Array (ATCA), observing in the same frequency band as ASKAP.  
One source has both a precise radio position (uncertainty 0.004 arcsec) measured with very long baseline interferometry and an optical position from DES.
We corrected a small residual offset in the DES image relative to the optical reference frame by cross matching stars in the DES images that had been cataloged by the Gaia mission \cite{GaiaDR2,online}.   
The positions agree with each other  within their uncertainties, confirming that the radio and optical frames are well aligned. 
We estimate the combination of statistical and systematic uncertainty in the burst's position to be $0.12$~arcsec in both right ascension and declination \cite{online}. 
The position of the burst is right ascension 21h44m25.255s $\pm$ 0.008s, declination $-$40$^{\circ}$54'00.1" $\pm 0.1"$ (equinox J2000). 

\section*{The burst host galaxy}

The sub-arcsecond localization for FRB~180924 allows us to uniquely identify the host by combining  public observations from the Dark Energy Survey \cite{DESDR1} with deeper images of the field we obtained with the Very Large Telescope (VLT), long-slit spectra with the Gemini South Telescope,  and integral-field spectra with the Keck-II telescope  and the VLT (see \cite{online} for details of instrumental setups).

Figure \ref{fig:host2} shows a deep VLT image of the field around the burst position.   The burst source is located $0.8 \pm 0.1$~arcsec  from the center of galaxy DES~J214425.25$-$405400.81 (galaxy A in  Fig. \ref{fig:host2}A)  cataloged by the DES \cite{DESDR1}.  
Keck observations establish the spectroscopic redshift of this galaxy to be $z=0.3214$, based on the ionized-Oxygen emission from diffuse gas in the galaxy  and  Calcium-absorption lines from its stellar component (Figure \ref{fig:host2}B).
The redshift was confirmed with  spectroscopic observations of the galaxy with Gemini-South (Figure \ref{fig:host2}C-D), which showed line emission from additional species at the same redshift, including  the first two (Hydrogen) Balmer transitions (H$\alpha$ and H$\beta$) and ionized Nitrogen \cite{online}.  
The deeper images obtained with the VLT show two other nearby objects which were also both detected in the integral-field spectra. 
There is  faint ionized-Oxygen emission from a dwarf galaxy, labeled galaxy B in Fig \ref{fig:host2},  at $z=0.384$  approximately 3~arcsec to the north west of the host and $\approx 3.6$~arcsec from the position of FRB~180924.
This corresponds to a projected distance of 19~kpc at the redshift of this galaxy.  
A third galaxy with a redshift $z=0.50055$ (galaxy C in Fig. \ref{fig:host2}) is located $3.5$~arcsec northeast of the  FRB position, at a projected distance of  21~kpc.  
We rule out association of the burst with these galaxies  with high confidence \cite{online}.

We  derive the properties of the host galaxy A by combining photometry from public surveys in optical \cite{DESDR1} and near-infrared wavelengths (from Wide-field Infrared Survey Explorer (WISE) $3.6~\mu$m and $4.5~\mu$m images) \cite{Wright10} with our optical imaging and spectroscopy (Figs. 2 and 3), using standard techniques  \cite{online}. 
The host properties are consistent with a massive lenticular or early-type spiral galaxy.  
The stellar population has a total mass of $2.2 \times 10^{10}$ ${\rm M}_\odot$ (${\rm M}_\odot$ is one solar mass), and is dominated by an old stellar population with an age $t_{age} > 4$\,Gyr. 
  The galaxy shows nebular emission lines with ratios consistent with gas excited by a harder spectrum than the ionizing flux of a star-forming population, characteristic of  low-ionization narrow emission-line region (LINER) galaxies \cite{yan12}.  
We demonstrate this by measuring the strength of forbidden transitions of  singly ionized nitrogen ([N~\textsc{ii}])  and doubly ionized Oxygen ([O~\textsc{iii}]), relative to, respectively,  H$\alpha$ and H$\beta$, and comparing to a well-studied sample of galaxies \cite{Baldwin81}, (Fig. \ref{fig:BPT}).
The host galaxy resides in the region of phase space occupied by LINER galaxies \cite{kewley01}. The galaxy also shows the presence of interstellar  dust, which is attenuating the optical-wavelength emission.  
The ratio of strength of H$\alpha$ to H$\beta$, combined with multi-band photometry suggest that there is internal  extinction  by dust within the galaxy (extinction $A_V \approx 1$ magnitudes, i.e., the optical $V$ band  is attenuated by a factor of $\approx 2.5$).
In principle, the measured Hydrogen 
emission lines can be used to constrain star formation in the galaxy. 
While the  data allow for a non-zero star formation rate,
 we report  an upper limit of $<2.0$ ${\rm M}_\odot$ per year, because we attribute a large fraction of  the dust-corrected
H$\alpha$ luminosity \cite{Kennicutt2012} to the LINER component.
The galaxy has a compact morphology described  by a S\'ersic profile \cite{Sersic63} with index $n = 2.0 \pm 0.2$ and an effective radius   of $2.79\pm 0.01$~kpc. 
 The burst is located exterior to $\approx$90\% of the galaxy's stellar light \cite{online}.   
 
We detect no radio-continuum emission from the burst location or anywhere within its host. We searched the host galaxy for  radio emission with ATCA in a continuous band from $4.5-8.5$\,GHz, at $1$ and $10$~days post-burst, and with ASKAP in a band from $1.1$ to $1.3$\,GHz $2$ days post burst. We set  $3\sigma$ flux-density limits on the emission of 20~$\mu$Jy  at a central frequency of 6.5 GHz and 450~$\mu$Jy at 1.3~GHz \cite{online}.

No repeated bursts were observed from this direction, before or after the burst was detected. We conducted sensitive searches with the Parkes radio telescope for a duration of 9~hr starting $8$~days post-burst and a further 2 hours, $23$ days post-burst \cite{online}.   
No pulses were found above a $10\sigma$ limit of $0.5 w_{\rm ms}^{-1/2}$~Jy\,ms for widths of $w_{\rm ms}$~ms. 
Likewise, no pulses were  found in $720$~hr of observations of the field as part of previous, less sensitive, single-antenna observations with ASKAP \cite{Shannon18} conducted between March $2017$ and February $2018$. 
These searches place $10\sigma$ limits on fluence of $25 w_{\rm ms}^{-1/2}$~Jy\,ms, for pulses of width $w_{\rm ms}$\,ms  \cite{online}.  
The burst was detected in a campaign in which the field was observed with a $10\sigma$ fluence limit of $5 w_{\rm ms}^{-1/2}$~Jy\,ms in a total observing time of $8.5$~hr.

\section*{Comparison to FRB~121102 and its host}

The properties of the burst and its host differ markedly to those of the repeating burst source FRB~121102 and its host galaxy. 
The host galaxy of FRB~180924 is a lenticular or early-type spiral with negligible or low rates of star formation.   
In contrast,  the host of the  FRB~121102 is a factor of $30$ less luminous, and is a low-mass, low-metallicity (low abundance of heavy elements), dwarf galaxy, with high star formation rate\cite{Bassa17}. Such dwarf galaxies are sites of high mass star formation, and frequent hosts of superluminous supernova and gamma-ray bursts\cite{Perley16}.
The two galaxies reside in completely different regions of the galaxy-type phase-space defined by their emission lines (Fig. \ref{fig:BPT}).

The burst source environments are also very different. 
FRB~121102  resides in a radio nebula containing highly magnetized plasma; its bursts have high rotation measures (RM$\sim 10^5$~rad\,m$^{-2}$, Ref. \cite{Michilli18}),  with the bursts showing a $10\%$ decrease over about a year \cite{Michilli18}.  
A large dispersion measure contribution is inferred from FRB~121102's host and local environment ($55$ to $225$~\pccm \cite{Tendulkar17,Bassa17}), indicating that it propagates through (and is likely embedded in) a dense, highly magnetized and dynamic plasma.
The  source of the repeating FRB~121102 is also co-located with a compact radio source with luminosity $1.8 \times 10^{22}$~W\,Hz$^{-1}$ \cite{Chatterjee17} at 6~GHz, while FRB~180924 shows no evidence for persistent associated radio emission at a limit $\sim 3$ times less luminous than the luminosity of the FRB~121102 compact source.

FRB~180924 has not been observed to repeat, despite extensive observations at low sensitivity with ASKAP and sensitive contemporaneous observations with the Parkes radio telescope.  
It is difficult to assess the statistical significance of the non repetitions from a single burst source. 
While the repetition rate of FRB~121102 is poorly characterized,  the activity appears to be clustered into week-month time scales \cite{Gourdji19,Oppermann18} followed by long periods of inactivity. Sensitive searches with the Parkes radio telescope shortly after FRB~180924 was discovered did not detect any further bursts on week-month timescales.

The differences between FRB~180924 and FRB~121102 -- the only other well localized burst source -- suggest  that either there could be two  different populations of burst progenitors, or that progenitors occur in diverse environments.
Models assuming a single progenitor class for bursts must reproduce the diversity in phenomenology and environments observed for burst sources.

\section*{Using the burst as an intergalactic and cosmological probe}

The dispersion and redshift of FRB~180924 can be used to test models of the free electron column density of the intergalactic medium (IGM). We model the dispersion to be the sum of components from the Milky Way's disk and halo, the intergalactic medium, and the burst's host galaxy.
Using models of the Milky Way, we infer a dispersion contribution from the disk \cite{NE2001} to be  $40$\,\pccm\ and the halo 
\cite{xyz19} to be $60$\,\pccm. 
A simple model of the intergalactic medium, based on the average baryon density and ionization fraction of the Universe \cite{xyz19} 
predicts the intergalactic component of the dispersion to be 
\dmcosmic~\pccm\, out to the redshift of the host. 
The sum of these components exceeds the dispersion of 
FRB~180924 by \dmexcess~\pccm\, 
without including any contribution  from the host galaxy interstellar medium and its halo. 
The errors in the Milky Way and halo components are expected to be small
($\sim 30$\,\pccm)
relative to the total dispersion budget \cite{xyz19}, 
so the main source of uncertainties in estimating the host dispersion contribution is the intergalactic medium  component. The latter depends on the distribution of foreground circumgalactic gas with respect to the associated dark matter halos (a process strongly influenced by galactic feedback) and 
sample variance along a given sight line. 

We use an IGM model that takes these uncertainties into account\cite{McQuinn14} to derive posterior probability distributions on the host electron densities under a range of assumed halo shapes \cite{online}. 
The mean host contribution to the dispersion inferred from these models,  corrected for host redshift is in the range $(30-81)$~\pccm,  with the 95\% upper limits ranging from $77-133$~\pccm. 
This indicates that dispersion of FRB~180924 is consistent with models of the IGM, provided the host contribution is 
much smaller than that found for FRB~121102.

There are two plausible locations for the burst temporal broadening: in the host galaxy or an intervening galaxy halo. 
It is unlikely that the burst is scattered by the diffuse extragalactic medium \cite{Macquart13}.
Similarly the temporal broadening in the Milky Way at high latitudes is predicted to be small at these frequencies (  $<1\,\mu$s, Ref. \cite{NE2001}).  
If the burst is scattered by the host galaxy, the medium has increased turbulence compared to the Milky Way  \cite{online}.  Substantially lower levels of turbulence would be required in an intervening galaxy halo to produce the measured scatter broadening, because turbulence near the midpoint between the source and observer produces relatively more broadening than if the same level of turbulence were at either end.  For a fixed turbulence strength, relative to an ISM line of sight a distance  $D_{\rm ISM}=10$~kpc in the host galaxy's ISM, the extragalactic line of sight has an enhancement in temporal broadening by a factor $D_{\rm IGM}/D_{\rm ISM} \sim 10^5$, for lines of sight of at a distance of $D_{\rm IGM}=10^6$~kpc in the IGM \cite{Macquart13}.  

The burst can be used to quantify the mean magnetization of the dispersing plasma along the line of sight. 
Assuming both uniform  magnetic field and electron densities along the line of sight, and using the excess Faraday rotation and dispersion of this burst, we set an upper limit on the magnetic field strength in the IGM parallel to the line of sight of $\lesssim 30 (1+z_{\rm EG})$~nG, where $z_{\rm EG}$ is the mean redshift of the magnetized plasma.  These constraints are similar to those found for previous bright bursts \cite{Ravi16}, and consistent with models of magnetization in extragalactic plasma \cite{Akahori16}.

Based on our sub-arcsecond localization of FRB 180924 to a galaxy at $z=0.3214$, we expect single-pulse fast radio bursts to be potential probes of the intergalactic medium at cosmological distances.
Firstly, the rate of detection of single event bursts is a factor  $>30$  greater than those that have been found to repeat, so we expect them to provide a larger statistical sample. 
Secondly, if the environment of FRB~180924 is representative, this population of bursts have relatively small uncertainties in estimating the density and magnetization  of the IGM out to large distances. 
Finally, if the hosts of other bursts are similarly luminous as the host of FRB~180924, identifying  hosts at high redshift will be easier than if bursts are exclusively hosted in dwarf galaxies \cite{Eftekhari17}, like the host galaxy of FRB~121102. 

\clearpage

\clearpage
\newpage

\begin{figure}[htpb!]
    \centering
    \begin{tabular}{c}
 \includegraphics[width=1.0\linewidth]{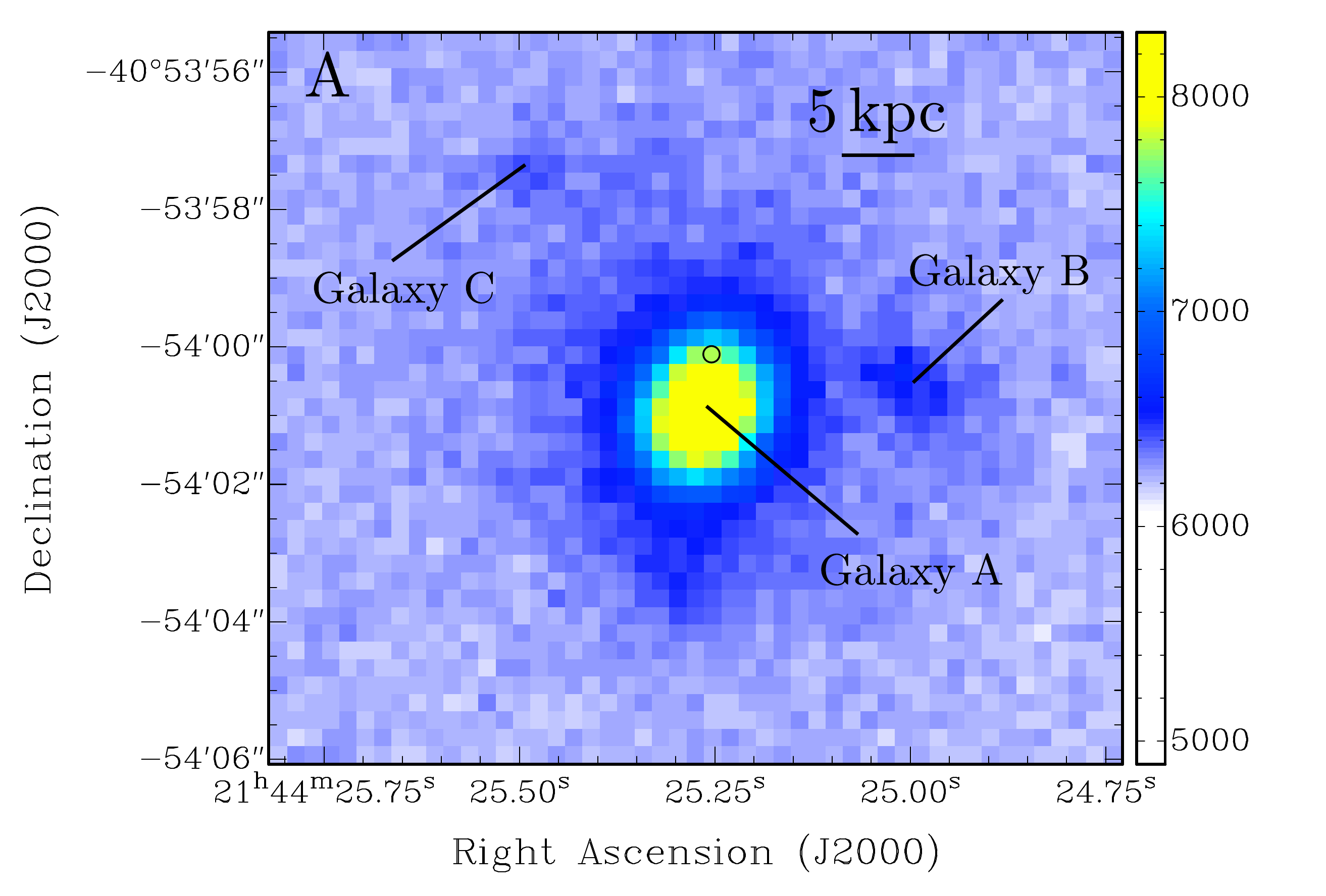} 
\end{tabular}
    \caption{Panel A: {\bf Host galaxy of FRB~180924.}  (A)   VLT/FORS2 $g^{\prime}$-band image showing the host galaxy of FRB~180924, labeled A. 
    The burst location uncertainty is shown by the black circle.  Two background faint background galaxies, labeled B and C, can be seen to the right and upper left are also visible (see supplementary text).  
     }
    \label{fig:host2}
\end{figure}

\clearpage
\newpage

\setcounter{figure}{1}

\begin{figure}[htpb!]
    \centering
    \begin{tabular}{c}
 \includegraphics[scale=0.5]{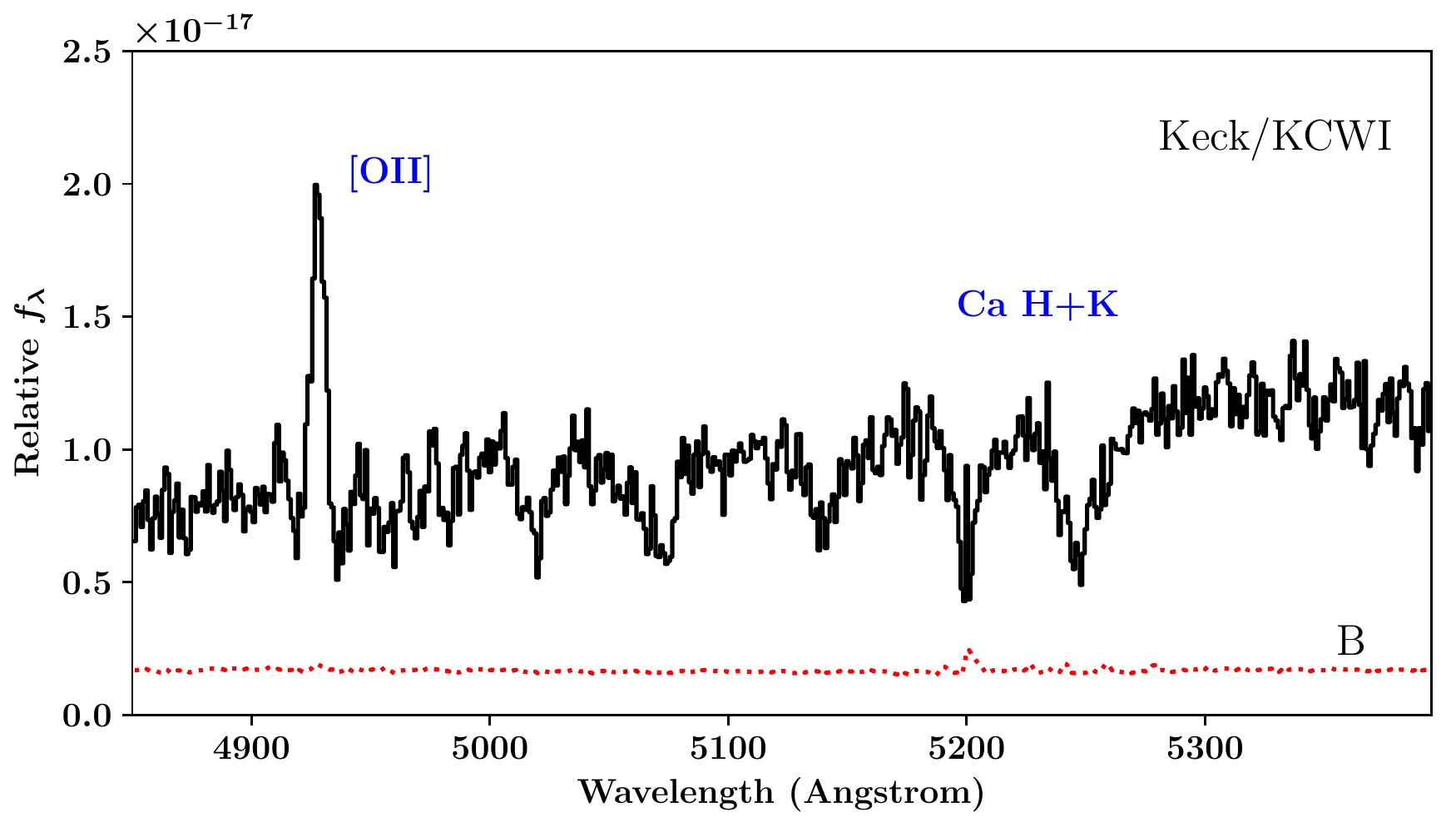}  \\ 
\includegraphics[scale=0.5]{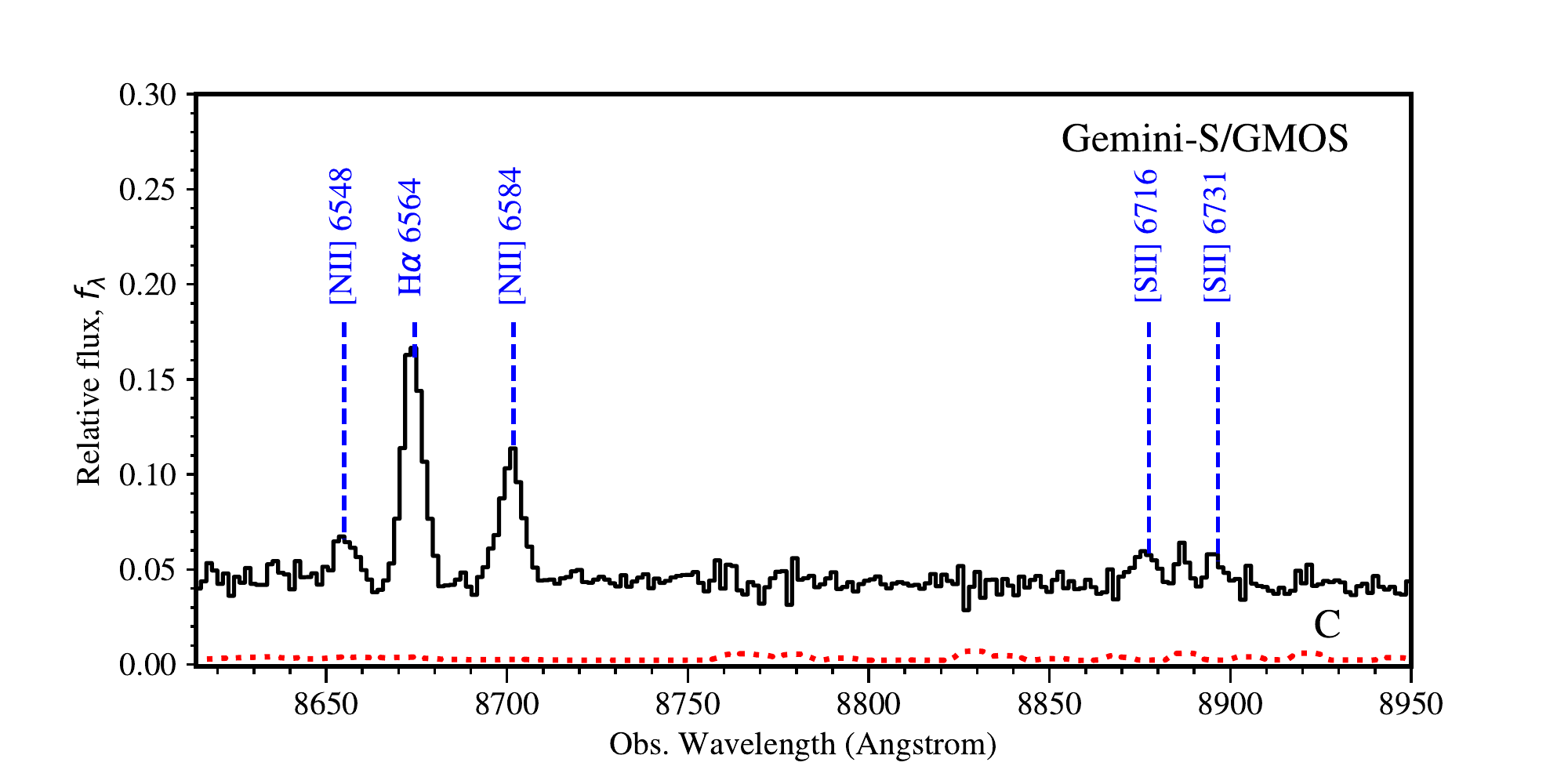}  \\
\includegraphics[scale=0.5]{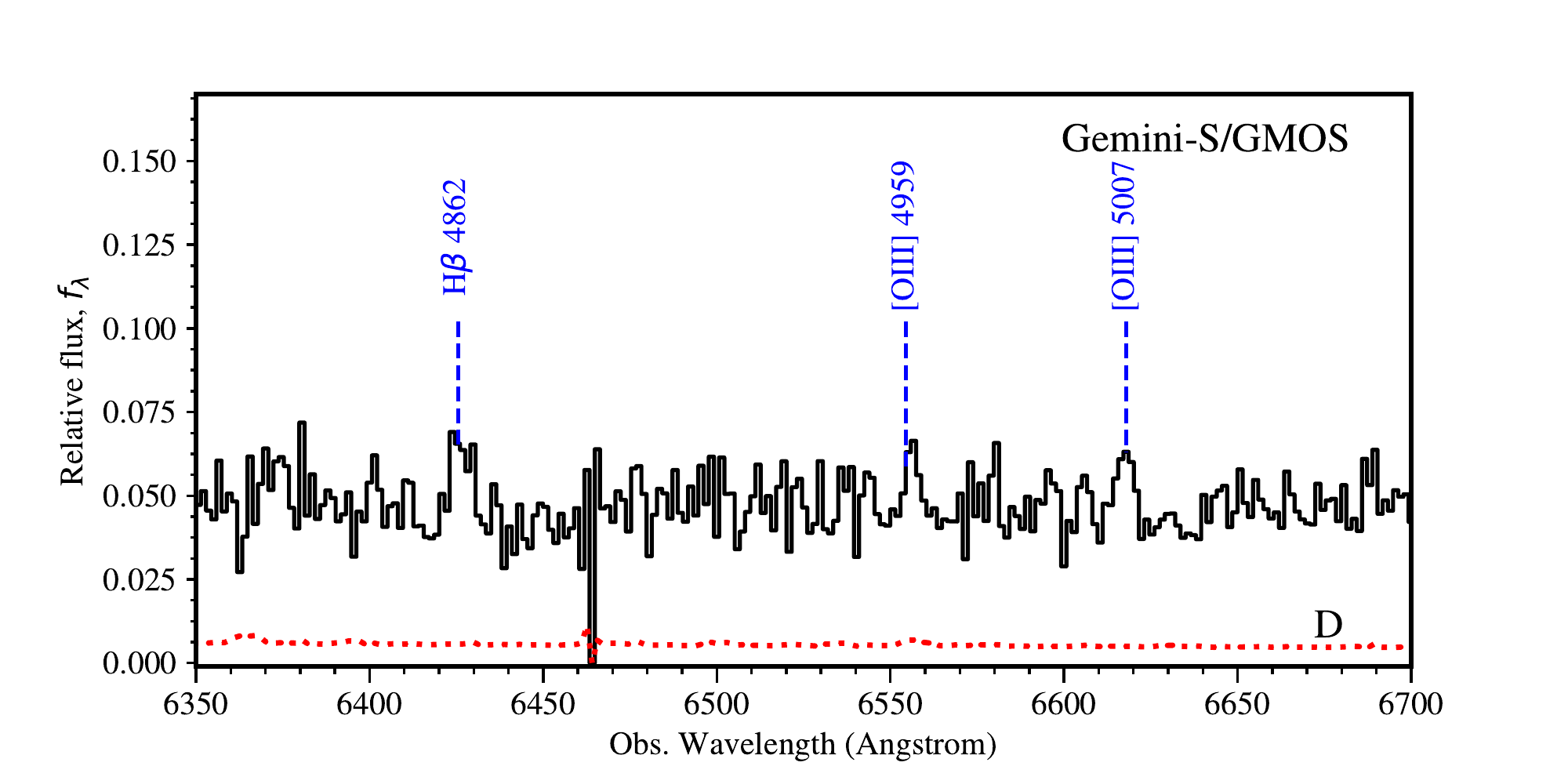} \\
\end{tabular}
    \caption{ Panels B-D:  {\bf Host galaxy of FRB~180924.}  
    (B)   Keck Cosmic Web Imager (KCWI) spectrum \cite{online} of the FRB~180924 host, showing the detection of forbidden-line ionized oxygen emission [O\textsc{ii}], and Calcium absorption 
    which set the FRB redshift $z=0.3214$. $f_\lambda$ is relative flux. 
    The oxygen emission is attributed to gas ionized by a hard ionizing spectrum.
    The absorption lines are stellar. 
    (C)  Section of Gemini Multi Object Spectrograph (GMOS) spectrum.  The spectrum shows (Hydrogen) Balmer line H$\alpha$, Nitrogen, and
    Sulfur emission at a redshift consistent with the 
    lines detected in the Keck spectrum. 
    (D)  Section of GMOS spectrum showing detections of the Balmer line H$\beta$ and forbidden line emission from  doubly ionized Oxygen ([O~\textsc{iii}]). 
    }
\end{figure}

\clearpage
\newpage

\begin{figure}[htpb!]
    \centering
\includegraphics[width=\linewidth]{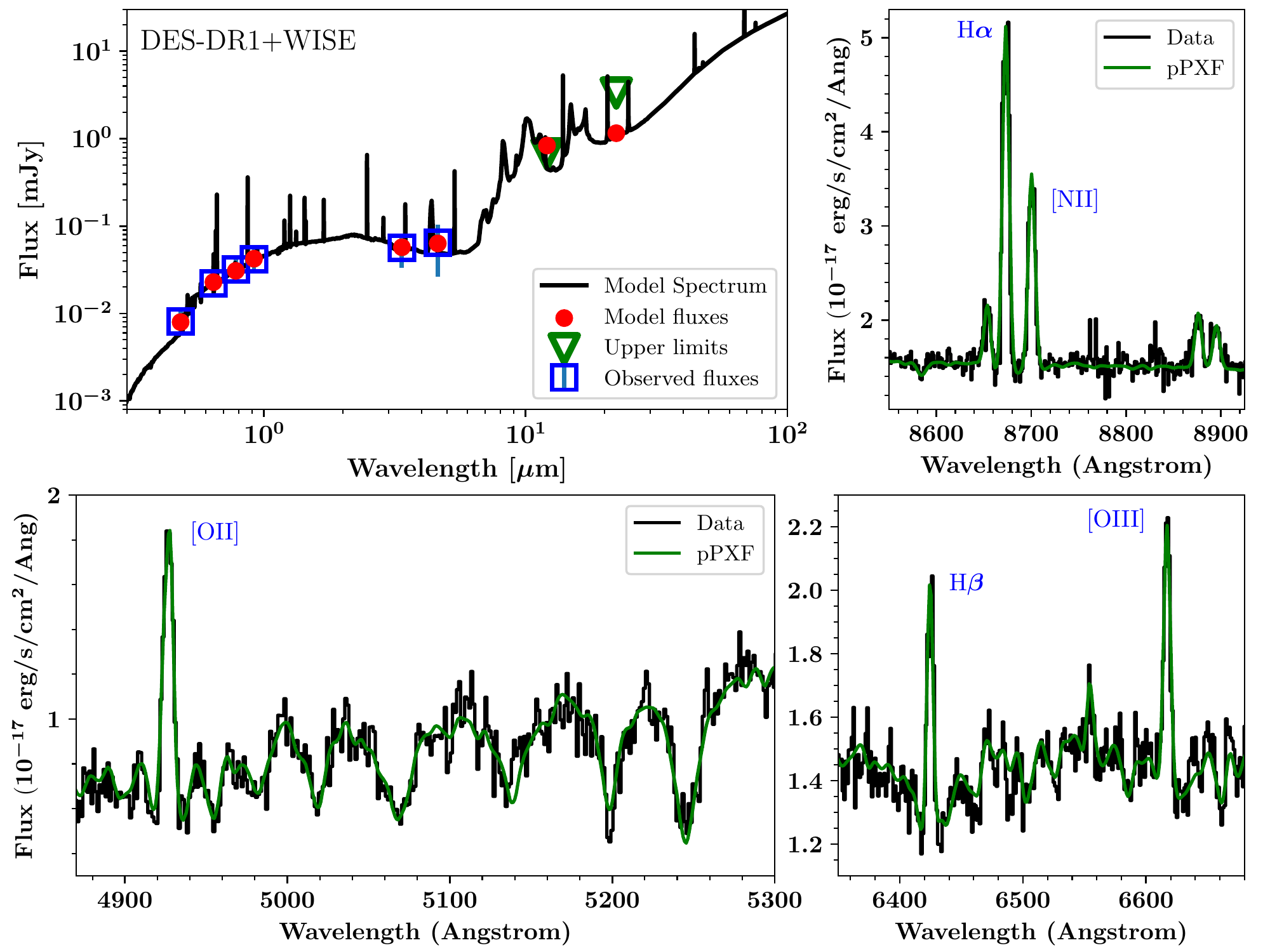}
    \caption{{\bf Spectrophotometric properties of FRB~180924 host galaxy.} 
    (A) photometric measurements of the host galaxy. 
    Spectral energy distribution (SED) modeling 
    yields an estimated stellar mass
    of $M_* = \stellarmass$ dominated by a modestly reddened,
    old stellar population ($t_{\rm age} > 4$\,Gyr).
    (B-D)
    Nebular line emission of the FRB~180924 host from the
    VLT/MUSE data cube.  Penalised Pixel Fitting (pPXF) models fitted to these data (green lines)  
    yield [O~\textsc{iii}]/H$\beta$ and [N~\textsc{ii}]/H$\alpha$ line-ratios
    characteristic of early-type LINER galaxies.   (B)  Spectrum includes  Hydrogen Balmer  line (H$\alpha$) and forbidden Nitrogen ([N~\textsc{ii}]).  (C)  Spectrum includes forbidden oxygen ([O~\textsc{ii}])  and  stellar absorption features.   (D)  Spectrum includes Hydrogen Balmer  line (H$\beta$) and forbidden oxygen ([O~\textsc{iii}]) transitions.    
    }
    \label{fig:host}
\end{figure}

\clearpage
\newpage

\begin{figure}[htbp!]
    \centering
   \includegraphics[width=0.8\linewidth]{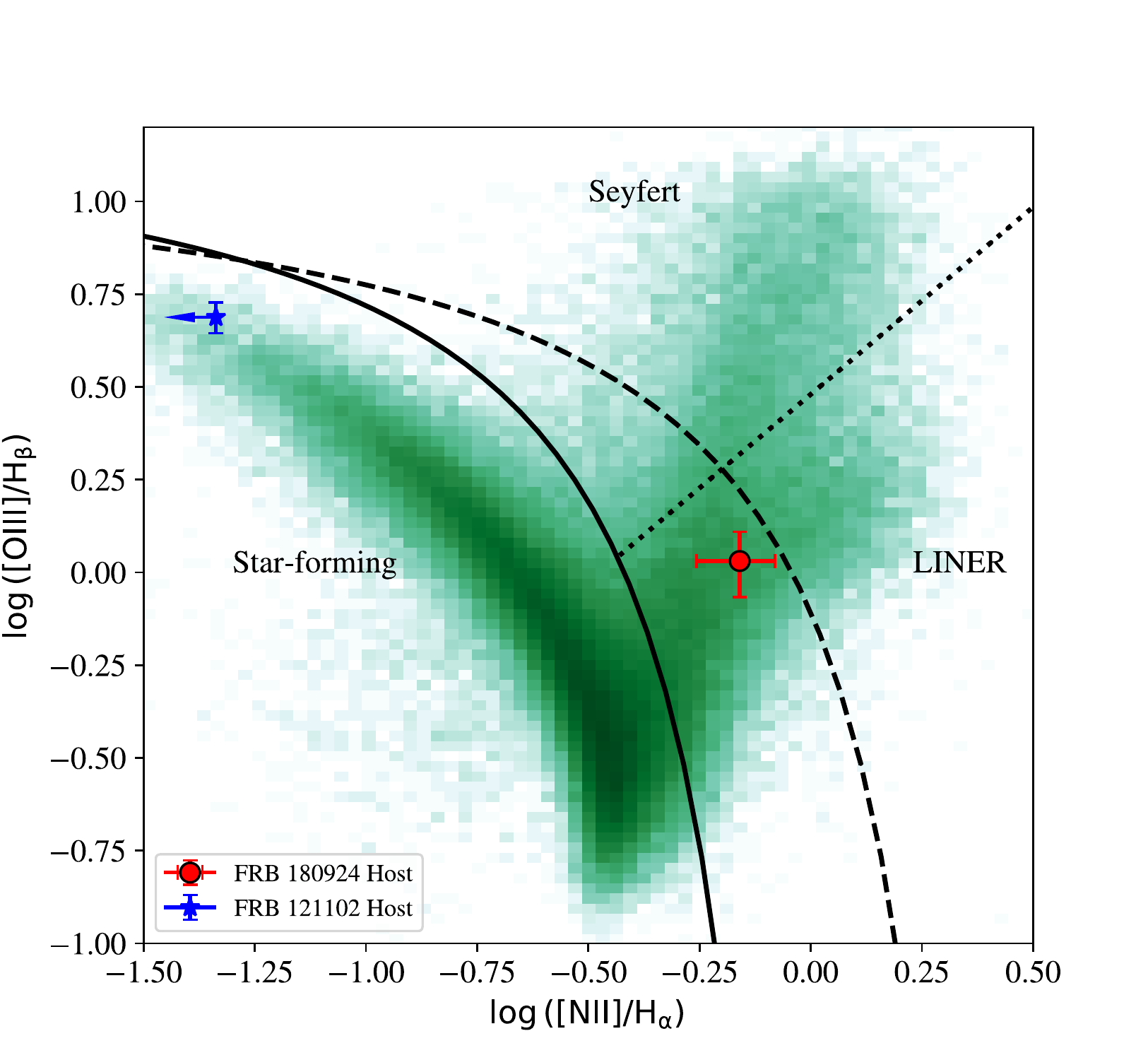}
    \caption{
    {\bf Diagnostic plot for characterizing emission line galaxies.}
        The green points show the distribution
    of $\approx 75,000$ nearby ($0.02 < z < 0.4$) emission-line galaxies from the Sloan Digital Sky survey, 
    restricted to have S/N~$>5$.  The intensity scaling is logarithmic
    to accentuate regions away from the dominant, star-forming
    locus.   
    Black lines separate the star-forming galaxies
    (solid; \cite{kewley01}) 
    from sources dominated
    by hard spectra (dashed; \cite{kauffmann03}),
    and the
    dotted line separates sources designated as Active Galactic Nuclei into either Seyfert
    or LINER galaxies \cite{fernandes10}.
    The host galaxy of FRB~180924 (red circle) is well offset from the star-formation
    locus and most consistent with LINER emission.
    In contrast,
    the host galaxy of the repeating FRB~121102 falls firmly 
    in the star-forming sequence (blue star).  
    Its [N~\textsc{ii}]/H$\alpha$ ratio is formally an upper limit).  
    } \label{fig:BPT}
\end{figure}

\clearpage
\begin{table}[]
\centering
\caption{ {\bf Properties of FRB~180924 and its host.} The fluence is derived from incoherent sum data. The implied isotropic energy density has been corrected to emission rest frame using a spectral index of -1.6\cite{Macquart18}.
The redshift inferred from the DM has large scatter is expected about the mean trend \cite{McQuinn14} as discussed in the main text and the supplementary information. The Milky-Way DM component was estimated from the NE2001 model \cite{NE2001}. }
\label{tab:burst_properties}
\begin{tabular}{|l r|}
\hline
FRB properties & \\
\hline
Dispersion Measure (DM)  &  $361.42 \pm 0.06$~pc~cm$^{-3}$\\
Arrival time at 1152~MHz & 2018-09-24 16:23:12.6265 UT \\
Fluence  & $16 \pm 1$ Jy\,ms \\
Pulse width & $1.30 \pm 0.09$~ms \\
Right Ascension (J2000) & 21h44m25.255s $\pm$ 0.008s    \\ 
Declination (J2000) & $-$40$^{\circ}$54'00.1" $\pm 0.1"$\\
Galactic Longitude  &  0.742467 deg \\
Galactic Latitude &  $-$49.414787 deg  \\
Incoherent Detection S/N & 21 \\
Image frequency-weighted S/N & 194 \\
Fractional linear polarization & $80 \pm 10$\% \\
Spectral modulation index & 0.80 \\
Decorrelation bandwidth & 8.5~MHz \\
Rotation Measure & $14\pm1$~rad~m$^{-2}$\\ \hline
Host-galaxy properties & \\
\hline
Redshift & $0.3214 \pm 0.0002$ \\
Right Ascension (J2000) & 21h44m25.25s\\
Declination (J2000) & $-$40d54m00.81s \\
$r$-band mag & $20.54 \pm 0.02$  \\
$g$-band mag & $21.62 \pm 0.03$ \\
$i$-band mag & $21.14 \pm 0.02$\\
($g-r$) mag & $1.08 \pm 0.04$ \\ 
($r-i$) mag & $0.41 \pm 0.03$ \\
WISE $3.6~\mu$m mag &  $16.9 \pm 0.1$  \\
WISE $4.5~\mu$m mag & $16.1 \pm 0.2$  \\
Radio continuum ($1.4$~GHz, $3 \sigma$) & $< 450$~$\mu$Jy \\
Radio continuum ($6.5$~GHz), $3 \sigma$) & $< 20.4$~$\mu$Jy \\
\hline
Inferred properties & \\
\hline
Implied FRB isotropic energy density & $6 \times 10^{31}$~erg\,Hz$^{-1}$ \\
Redshift inferred from DM \cite{Inoue04} & 0.34 \\
DM Milky-Way Disk  & 40.5 \pccm \\
DM Milky-Way Halo  & 30 \pccm \\ 
\hline
\end{tabular}

\end{table}

\newpage

\clearpage



\section*{Affiliations}
\normalsize{ $^{1}$ Commonwealth Science and Industrial Research Organisation,  Australia Telescope National Facility, P.O. Box 76, Epping, NSW, 1710 Australia}\\
\normalsize{$^{2}$ Centre for Astrophysics and Supercomputing, Swinburne University of Technology, Hawthorn, VIC, 3122 Australia}\\
\normalsize{$^{3}$ International Centre for Radio Astronomy Research, Curtin University, Bentley WA 6102, Australia}\\
\normalsize{$^{4}$ University of California Observatories-Lick Observatory, University of California, 1156 High Street, Santa Cruz, CA 95064, USA}\\
\normalsize{$^5$Kavli Institute for the Physics and Mathematics of the Universe,
5-1-5 Kashiwanoha, Kashiwa, 277-8583, Japan} \\
\normalsize{$^{6}$ Instituto de F\'isica, Pontificia Universidad Cat\'olica de Valpara\'iso, Casilla 4059, Valpara\'iso, Chile}\\
\normalsize{$^7$Department of Physics and Astronomy, Macquarie University, NSW 2109, Australia } \\
\normalsize{$^{8}$ Sydney Institute for Astronomy, School of Physics, University of Sydney, Sydney, NSW 2006, Australia}\\
\normalsize{$^{9}$ Astronomy Department, University of Washington, Seattle, WA 98195, USA}\\
\normalsize{$^{10}$  Physics Department, California Polytechnic State University, 1 Grand Avenue, San Luis Obispo, CA 93407, USA}\\
\normalsize{$^{11}$ Nikhef, Science Park, Amsterdam, the Netherlands} \\
\normalsize{$^{12}$ International Centre for Radio Astronomy Research, The University of Western Australia, M468, 35 Stirling Highway, Crawley, Perth, WA 6009, Australia}\\
\normalsize{$^{13}$ Space Science Division, Naval Research Laboratory, Washington, DC 20375-5352, USA}\\
\normalsize{$^{14}$ W. M. Keck Observatory 65-1120 Mamalahoa Hwy, Waimea, HI 96743, USA}\\
\normalsize{$^{15}$ Department of Physics and Astronomy, University of California at Los Angeles, Los Angeles, CA 90095-1547, USA}\\
\normalsize{$^{16}$ Department of Physics and Astronomy, 4129 Frederick Reines Hall, University of California, Irvine, CA, 92697, USA}\\
\normalsize{$^{17}$ Tim Cornwell Consulting, 17 Elgan Crescent, Sandbach CW11 1LD, United Kingdom} \\
\normalsize{$^{18}$ Inter-University Centre for Astronomy and Astrophysics, Post Bag 4, Ganeshkhind, Pune 411 007, India}\\
\normalsize{$^{19}$ Australian National University, Research School of Astronomy \& Astrophysics, Canberra ACT 2611 Australia}\\
\normalsize{$^{20}$ Western Sydney University, Locked Bag 1797, Penrith South, NSW 2751, Australia}\\
\normalsize{$^{21}$ ASTRON, The Netherlands Institute for Radio Astronomy, Postbus 2, 7990 AA, Dwingeloo, The Netherlands}\\

\section*{Acknowledgments}
We are grateful to the Australia Telescope National Facilities (ATNF) operations team and the Murchison Radio-astronomy observatory staff for supporting these ASKAP searches, and the ATNF steering committee for allocating time for these observations. 
We thank Luca Rizzi for his support with the Keck Cosmic Web Imager instrument.
 K.W.B. thanks Y. Shalem for discussions.
 We acknowledge the Wajarri Yamatji as the traditional owners of the Murchison Radio-astronomy Observatory site. 
The authors recognize and acknowledge the very significant cultural role and reverence that the summit of Mauna Kea has always had within the indigenous Hawaiian community.
We are most fortunate to have the opportunity to conduct observations from this mountain.

  Full facility acknowledgements are provided in the supplementary materials.
 
 \subsection*{Funding}
 K.W.B., J.P.M., and R.M.S. acknowledge Australian Research Council (ARC) grant DP180100857.
 A.T.D. is the recipient of an ARC Future Fellowship (FT150100415).
 S.O. and R.M.S. acknowledge support through ARC grant FL150100148. 
 R.M.S. also acknowledges support through ARC grant CE170100004. N.T. acknowledges support from PUCV research funding 039.333/2018.
 Work at the Naval Research Laboratory is supported by NASA.
 
\subsection*{Author contributions}
K.W.B. built the search and voltage capture software. A.T.D., C.P., C.K.D., S.B., W.R.A., and M.B., designed, built and conducted the correlation, calibration and imaging software to localize FRB~180924. R.M.S. drafted the manuscript, led the observing, and interpreted radio band polarization data. J.P.M., J.X.P., N.T., S.D.R. E.M.S., S.S., M.M., F.O.N-H.,  V.N.B., J.B, J.O'M. T.T., V.U. obtained, reduced and interpreted optical observations.  M.K., E.L., E.K.M., S.O., H.Q., W.F, C.F., C.W.J., R.D., obtained, reduced, and interpreted radio data.  R.D.E, T.J.B., D.C.-J.B, R.J.B, A.B., J.D.B., A.P.C. F.R.C., T.C., N.G., D.B.H., M.K., B.S.K., A.M., N.M.-G. S.N., R.P.N., M.P.,R.-Y. Q., J.R., D.N.R., T.W.S., M.A.V., C.D.W. designed, built, and commissioned ASKAP. 

\subsection*{Competing Interests}
The authors declare no competing interests.

\subsection*{Data and materials availability:}

Based on observations collected at the European Southern Observatory under ESO programs 0102.A-0450(A) and 2102.A-5005(A) (PI: Macquart), available from \url{http://archive.eso.org/}. Observations obtained at Gemini Observatory under program GS-2018B-Q-133 (PI: Tejos) can be retrieved from \url{https://archive.gemini.edu/}. Observations from  the Australia Telescope Compact Array under program C3211 (PI: Shannon), and the Parkes radio telescope under program P958 (PI: Shannon) can be retrieved from \url{https://atoa.atnf.csiro.au/}. Keck observations were obtained under project ID U176 2018B and can be retrieved from \url{https://www2.keck.hawaii.edu/koa/}

Further datasets used in this paper are available from the CSIRO Data Access Portal at \url{https://doi.org/10.25919/5d09d22f2c004} : seven visibility data sets used to calibrate and determine the localisation of FRB~180924, the ATCA image used for astrometry, calibrated optical images and spectra, and radio images.

Data reduction scripts and code written by the co-authors for this project are available from the {\sc craft} git repository \url{https://bitbucket.csiro.au/scm/craf/craft.git} and the {\sc psrvlbireduce} repository \url{https://github.com/dingswin/psrvlbireduce}.

\subsection*{Supplementary Materials:}

Materials and Methods\\
Supplementary Text\\
Figures S1 to S7\\
Tables S1 to S9\\
References \textit{(42-97)}


\renewcommand\thesection{S\arabic{section}} 
\setcounter{section}{0} 

\renewcommand\thetable{S\arabic{table}} 
\setcounter{table}{0} 

\renewcommand\thefigure{S\arabic{figure}}
\setcounter{figure}{0}

\renewcommand\theequation{S\arabic{equation}}
\setcounter{equation}{0}


\begin{center}
{\LARGE
Supplementary Materials for\\}
{\large
 A single fast radio burst  localized to a  massive galaxy at cosmological distance}
\end{center}
\pagenumbering{gobble}

\noindent 
K.~W. Bannister, 
A.~T.~Deller, 
C.~Phillips, 
J.-P.~Macquart,
 J.~X.~Prochaska,
 N.~Tejos, 
  S.~D.~Ryder,
  E.~M.~Sadler,
  R.~M.~Shannon,
   S.~Simha, 
   C.~K.~Day,
    M.~McQuinn,
     F.~O.~North-Hickey
      S.~Bhandari
W.~R. Arcus, 
V.~N.~Bennert,
 J.~Burchett,
M.~Bouwhuis, 
R.~Dodson, 
R.~D.~Ekers, 
W.~Farah, 
C.~Flynn, 
C.~W. James, 
M.~Kerr, 
E.~Lenc, 
E.~K.~Mahony, 
J.~O'Meara, 
S.~Os{\l}owski,
H.~Qiu, 
T.~Treu, 
V.~U, 
T.~J. Bateman, 
D.~C.-J.~Bock, 
R.~J.~Bolton, 
A.~Brown, 
J.~D.~Bunton,
A.~P. Chippendale, 
F.~R. Cooray, 
T.~Cornwell,
N.~Gupta, 
D.~B.~Hayman, 
M.~Kesteven, 
B.~S.~Koribalski, 
A.~MacLeod, 
N.~M.~McClure-Griffiths, 
S.~Neuhold,
R.~P.~Norris,
M.~A.~Pilawa,
R.-Y.~Qiao, 
J.~Reynolds,
D.~N. Roxby, 
T.~W. Shimwell,
M.~A.~Voronkov, 
C.~D. Wilson

\noindent
{\bf This PDF file includes}\\
Materials and Methods\\
Supplementary Text\\
Figures S1-S7 \\
Tables S1-S9 \\

\newpage
\pagenumbering{arabic}
    \setcounter{page}{1}

\section{Materials and Methods}

\subsection{Follow-up observations} 

A summary of the observations used in this analysis is displayed in Table~\ref{tab:ObsSummary}.  

\subsection{ASKAP description and configuration for FRB searching}
The basic components of the ASKAP, and its FRB detection pipeline are described previously \cite{Bannister17}. 
We have used a similar setup.
Briefly, ASKAP comprises $36$, $12$-m antennas, each of which has a phased-array feed (PAF) at its focus. For this work, $24$ antennas were used.
Each PAF has 94 dual linear polarization receivers sensitive to the frequency range 700 to 1800 MHz. The sampled electric field from each element is digitized and channelized to 1 MHz resolution and sent to digital beamformers.
For this work, the system was configured to a center frequency of $1320$~MHz and bandwidth of $336$~MHz. The beamformers produce $36$ configurable dual-polarization beams by applying independent complex-valued weights to each of the ports, at $1$~MHz intervals.   
We used a $6\times6$ hexagonal closed packed beam configuration (see \cite{Shannon18}, their figure S1). The beamformers produce fast-dump autocorrelation spectra by squaring and averaging the beamformed voltages over a 864 microsecond window \cite{Clarke14}. These spectra are streamed over Ethernet to a dedicated fast transient searching node and processed on a single graphics processing unit (GPU). Simultaneously, the beamformers store the beamformed voltages in a configurable circular buffer. The buffers were configured to save all 36 dual polarization beams, with the voltages quantized to 4-bit  complex-valued integers with a 3.1-second buffer length. The buffer can be stopped and downloaded to disk, either manually, or under control of the detection pipeline.

\subsection{Search Pipeline}

We search the data using a custom GPU-based detection pipeline termed {\tt FREDDA} (Fast Real-time Engine for DeDispersing Amplitudes \cite{fredda}) which is optimized for ASKAP processing and capable of detecting bursts with low latency. Firstly, the pipeline normalizes the incoming spectra to have zero mean and unit variance in blocks of 1024 time samples, by measuring the mean and variance over the previous block and subtracting the mean and dividing by the variance independently for each channel, polarization and antenna.  The pipeline zero-weights channels where the mean or variance change by more than $20\%$ in adjacent blocks, or with excess kurtosis greater than 30. These steps effectively flag strong radio-frequency interference (RFI) from satellite transmitters.  The pipeline also flags individual integrations that are a factor of ten greater than the standard deviation. 
This flagging could excise exceptionally bright bursts, and would flag broad band bursts with  S/N above $10\sqrt{N_{\rm ant} N_{\rm chan}}\approx 900$. 
This effectively flags RFI from aircraft transponders. After rescaling and flagging, the pipeline sums across both polarizations and  $N_{\rm ant}$ antennas to produce a single dynamic spectrum for all available antennas. 
This incoherent sum results in an improvement in sensitivity by a factor of $(2 N_{\rm ant})^{1/2}$ than a single polarization from an antenna (if the signal is unpolarized). 
The pipeline then performs a Fast Dispersion Measure Transform \cite{Zackay17} - searching $2048$ DM trials - equivalent to a DM range of 0-1285~\pccm with a DM resolution of 0.63~\pccm.
Finally, the pipeline performs a boxcar convolution of all of the dedispersed time series with widths of 1 to 32 samples. Candidates with  $S/N > 10$ are transmitted as UDP packets to a second computer program that triggers the voltage buffers. This second program declares a trigger when a candidate with $S/N > 10$ is detected with a width $w < 8$ integrations.    This width was chosen to reduce the number of false triggers.  When these criteria are met, it stops the circular buffer and downloads the beamformer voltage buffers for the desired beam; once a download is completed, data flow into the circular buffer is restored and the system is available for a new trigger.

\subsection{Interferometry of ASKAP Voltage Data}

To robustly confirm the location of the FRB, two teams independently correlated, calibrated, and imaged the burst using entirely separate software pipelines, without sharing results before each had obtained a position.

One team used a custom correlator that was originally created for commissioning purposes, which read the antenna voltage data in raw format, correlated it, and processed the resulting data in {\sc Miriad} \cite{Sault95}. The second team converted the stored antenna voltage data into a format readable by the  DiFX software correlator \cite{Deller11} and correlated the data using this package, using incoherent dedispersion to align the FRB emission in frequency.  The resulting files (in FITS-IDI format) were calibrated with standard techniques using AIPS \cite{Greisen03};  the calibrated data was then imaged using CASA version 5.3.0-143 \cite{CASA} .

Both teams obtained nearly identical results, with the FRB position agreeing to within $1\sigma$. The dataset produced by the DiFX correlator (which has been verified to sub-milliarcsecond levels of accuracy when correlating very long baseline radio interferometers \cite{Deller19}, and hence is expected to introduce negligible systematic position offsets) was adopted as our fiducial result.  Both processing pipelines were tested with a long (10-hr) observation of a strong continuum source to show a stable phase and amplitude response with time. 

In order to calibrate the FRB visibility datasets, 5 hr after the detection, we observed a  bright compact  ($< 10$~mas) radio galaxy PKS~0407$-$658, which was placed at the center of the beam in which the burst was detected. The voltages obtained were processed identically as those for the FRB.  The Vela pulsar (PSR~J0834$-$4510) was observed 12 hr after the FRB was detected and a voltage download similarly triggered and correlated.

A total of seven visibility datasets were produced using DiFX and the three sets of voltage data, listed below.  Each included full-polarization products and a frequency resolution of $9.26$~kHz, averaged post-correlation to $148$ kHz resolution.

\begin{itemize}
\item PKS~0407$-$658 data, correlation centered at PKS~0407$-$658 (R.A. 04h08m20.38s, Decl. $-$65d45'09.08"), time resolution 1.3824 s, total integration time 3.1 s (the {\em  calibrator} dataset).
\item FRB~180924 data, correlation centered at the ASKAP PAF beam center position (R.A. 21h45m17.83s, Decl. $-$41d03'34.67"), time resolution 1.3824 s, total integration time 3.1 s (the {\em field} dataset).
\item FRB~180924 data, correlation centered at approximate FRB position (R.A. 21h44m25.2943s, Decl. $-$40d53'59.9959"), time resolution 1~ms, total integration time 1ms, dedispersed with DM 361.53~\pccm~(the {\em FRB position} dataset).
\item FRB~180924 data, correlation centered on approximate FRB position (R.A. 21h44m25.2943s, Decl. $-$40d53'59.9959"), time resolution $0.2$~ms, total integration time 3~ms, de-dispersed with DM 361.53~\pccm~(the {\em FRB structure} dataset).  The small offset (0.2 arcsec)  between the correlated phase center and the burst position does not affect astrometric precision or accuracy. 
\item FRB~180924 data, correlation centered at approximate FRB position (R.A. 21h44m25.2943s, Decl. $-$40d53'59.9959"), time resolution 33~ms, total integration time 30~ms (centered on the FRB time, but excluding 3~ms centered on the FRB itself), de-dispersed with DM 361.53\pccm~(the {\em  FRB RFI subtraction} dataset).
\item Vela data, correlation centered at the location of the Vela pulsar (R.A. 08h35m20.61149s, Decl. $-$45d10'34.8751"), time resolution $1.3824$~s, total integration time 3.1~s, de-dispersed with DM 67.99~\pccm, using pulsar gating to select only 0.9~ms from  every rotation of the pulsar when the emission is brightest (the {\em Vela} dataset).
\item Vela data, correlation centered at the location of the Vela pulsar  (R.A. 08h35m20.61149s, Decl. $-$45d10'34.8751"), time resolution 1.3824~s, integration time 3.1~s, de-dispersed with DM 67.99~\pccm, using pulsar gating to select 30~ms of data per Vela rotation from times adjacent to but not on the Vela pulse (the {\em Vela RFI subtraction} dataset)
\end{itemize}

By inspecting the PKS~0407$-$658 data in the calibrator dataset, we identified and solved for residual, unmodeled antenna-based delays.  Once identified, these delays were removed from the visibility datasets by re-correlating all datasets while applying the necessary delay corrections in the correlator-delay model.

Calibration solutions were derived from the calibrator data set as follows.  Due to strong radio frequency interference, approximately 100 MHz of bandwidth concentrated towards the lowest frequencies of the observed band was flagged prior to the derivation of any solutions.  Residual antenna-based delay and phase errors were derived using the {\sc AIPS} task {\tt FRING}.  The flux density scale was set using the tasks {\tt SETJY} and {\tt CALIB}, assuming PKS~0407$-$658 to be a 9.5~Jy, flat spectrum, unpolarized source.  Finally, frequency-dependent amplitude and phase corrections accounting for the instrumental bandpass were derived using the task {\tt CPASS} to interpolate over the flagged regions of spectrum.  The resulting corrections were transferred from the calibrator dataset to all 6 other datasets and applied. 

First, we imaged the FRB to obtain its position.  To mitigate the effects of radio frequency interference, we subtracted a scaled version of the {\em FRB RFI subtraction} dataset from the {\em FRB position dataset}. Since the effect of the RFI on the visibilities is approximately constant on millisecond timescales, this effectively removes the RFI (and constant background celestial sources) from the FRB position dataset, leaving only signals that vary on millisecond timescales - i.e. the FRB itself.  A custom ParselTongue script \cite{Kettenis06}  was used to manipulate the visibilities directly in {\sc AIPS} to perform this subtraction (the {\sc uvsubScaled.py} task in the {\sc psrvlbireduce} repository).  We imaged this RFI subtracted dataset in CASA, initially using the task {\sc clean} in widefield multifrequency synthesis mode with natural weighting, producing a $2048\times2048$ pixel image with 3~arcsec pixels.  Once the source location was identified, we re-imaged the data producing a smaller $128\times128$ pixel image with a 1~arcsec pixel centered on the FRB.  We estimated the position and flux density of the FRB, and the corresponding uncertainties, using the {\sc AIPS} task {\tt JMFIT}.

Since the FRB is observed to be highly polarized and predominantly detected in only one linearly polarized receptor (Y),  we extracted positions using only the  YY polarization image, which has the highest signal to noise ratio.  
A simple combination of the entire 336 MHz bandwidth yielded a detection significance of $70 \sigma$. 
However, this is not an optimum approach for a source whose amplitude varies substantially across the observing band such as FRB~180924.  
To account for this and produce the highest possible S/N on the FRB, we first produced an image cube with 4 MHz resolution and used this to estimate the FRB flux density ratio (the ratio of instantaneous flux density to frequency-averaged flux density) as a function of frequency.  
We then used this estimate to re-weight and re-scale the FRB data with the {\sc uvfrbwt} task (see the {\sc CRAFT} repository), dividing the visibility amplitudes by the FRB flux density ratio and multiplying the visibility weights by the FRB flux density ratio squared.  This has the effect of normalizing the FRB amplitude to a constant value across the band while up-weighting the regions where the FRB is bright and down-weighting regions where it is faint.  

Using this rescaled and reweighted dataset, we obtained a detection of significance 184$\sigma$, with position (J2000) of R.A. 21h44m25.255s $\pm$ 0.003s and Decl. $-$40d54'00.10" $\pm$ 0.04", where the uncertainties are the statistical values reported by {\tt JMFIT} only.  
We find that the position of the source changes by a root-mean-square value of 70~mas in both R.A. and Decl. when we change the weighting used in the imaging, which exceeds the statistical uncertainty from {\tt JMFIT}.  We therefore use the larger 70~mas value as an estimate of the astrometric uncertainty for the burst position.

As a cross-check, we imaged only the upper regions of the frequency spectrum using the non-RFI subtracted FRB position dataset and obtained a consistent position with a larger uncertainty due to the lower signal-to-noise ratio. We also extracted the positions from each plane of the 4 MHz resolution image cube, confirming that they gave results consistent with the best-fitting position as a function of frequency.

To analyze the polarization properties of the burst, we produced a full Stokes image cube using the RFI-subtracted FRB position dataset, with a frequency resolution of $0.167$~MHz.  We used the {\sc casaviewer} tool to extract the mean pixel value within a small window centered on the best fit FRB position in each image slice, thereby obtaining the FRB spectrum in each Stokes parameter at $0.167$~MHz spectral resolution. 
We repeated the polarization analysis and positional analysis described above with the Vela datasets, performing RFI subtraction and imaging in an analogous manner.
To determine if there is an offset between the YY and I frames, we compared the positions of continuum sources in the field derived from the YY and Stokes-$I$ images.  We found the positions were consistent to within measurement uncertainty.

Finally, we imaged the {\em FRB field} dataset using CASA to determine the location of radio continuum sources that could be used to register the FRB position derived above in the International Celestial Reference Frame (ICRF3).

Since no RFI subtraction could be performed with this dataset, regions of strong RFI (as noted above, concentrated in the lower half of the band, and predominantly due to satellite transmitters) were identified and flagged prior to imaging, leaving 205 MHz of usable bandwidth.  We made use of the {\sc CASA} task  {\tt TCLEAN}  task using multi-scale multi-frequency synthesis and two Taylor terms to account for real or instrumental spectral index effects. Initially, a Stokes-$I$ image was made using natural weighting. 
Six continuum sources were detected above a S/N of 10, with 3 being sufficiently compact and bright to be useful for astrometry.
The positions of each ASKAP source were extracted using the {\sc AIPS} task {\tt JMFIT} (see Table \ref{tab:field_sources}).

In addition to the multi-frequency synthesis image spanning the full ASKAP bandwidth, we produced a Stokes-I image cube with 4-MHz resolution, yielding 58 (out of 83) 4~MHz frequency channels that could be used for fitting, and which were unaffected by RFI. Extracting positions for the brightest source, PKS~2144$-$418 from each channel of the cube showed no significant deviations as a function of frequency across the band. Stokes $Q$, $U$, and $V$ image cubes were also produced along with multi-frequency synthesis Stokes $Q$, $U$, and $V$ images to determine the level of polarization leakage, yielding no significant detections in the polarized Stokes parameters and a typical $2\sigma$ detection upper limit leakage range for each of 10-15\%.

After imaging the FRB field, we used the {\tt GAINCAL} task in {\sc CASA} to derive phase-only self-calibration solutions, using the ASKAP model of the field sources. A single solution was derived per antenna, spanning the entire integration time and bandwidth. The phase corrections derived in this manner were small (averaging a few degrees) and consistent with the noise expected given the ASKAP sensitivity and the brightness of the sources in the field.  Application of these self-calibration solutions did not substantially improve the dynamic range of the image, and so the self-calibration solutions were not applied to either this dataset or any of the FRB datasets.  

\subsection{Radio Astrometry}

To tie the ASKAP positions to the International Celestial Reference Frame (ICRF3) we observed the three bright continuum sources detected with ASKAP (PKS~2144$-$418,\break SUMSS~J214421$-$412640 and SUMSS~J214438$-$411835, see Table \ref{tab:field_sources}) with the Australia Telescope Compact Array (ATCA) on 2018 October 5. The observations were carried out from 1.1-3.1 GHz in the 6A configuration (which has baselines as long as 6 km) to maximise overlap with the ASKAP frequency band and angular resolution. Four ATCA phase calibrators (PKS~2106$-$413, PKS~2211$-$388, PKS~2052$-$474 and PKS~2054$-$377) were interspersed throughout the observations, spending 2~mins integration time on each of the calibrators and 4 mins on the ASKAP-detected sources. 

This process was repeated for a total of 11~hrs to maximize coverage of the u-v plane. The four phase calibrators used were selected to have very small phase closure defects on 6-km baselines and small position uncertainties. 

A significant amount of the band between 1.1-2.1 GHz was affected by RFI so only the 2.1-3.1 GHz band was used in the data reduction. The data was flagged, calibrated and imaged using standard {\sc Miriad} tasks. We used robust=$-1$ (close to uniform) resulting in beam sizes ranging from $3.4-3.8\times 2.4$~arcsec. Positions and associated errors were extracted using {\tt IMFIT} and fitting a point source. 

We investigated the ATCA astrometric uncertainties by calibrating using each phase calibrator separately, and applying the calibration solution to the remaining three calibrators and imaging and extracting the positions as described above. We found large errors in PKS~2211$-$388 which we suspect is due to structure in the source. The position extracted from the images agreed with the cataloged position within 100 mas in all other cases, with the measurement uncertainties typically 10 mas.
We calibrated the target sources based on the phase calibrator PKS~2052$-$474 only, which is a  {\em Defining} source for the ICRF3 frame. 

\begin{figure}[htpb!]
    \centering
    \includegraphics[width=\linewidth]{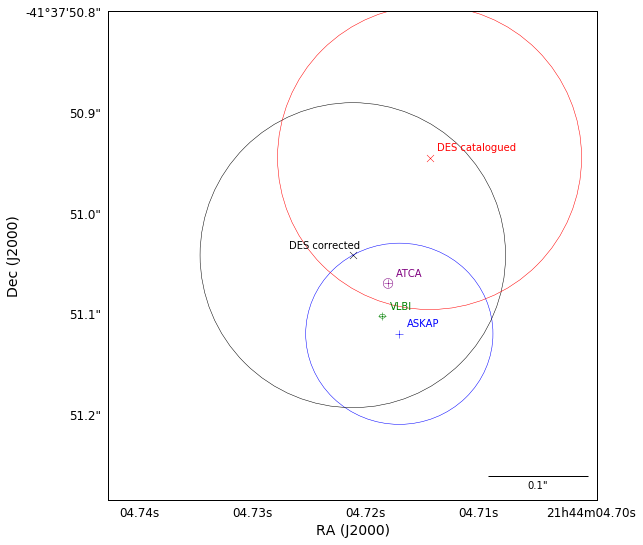}
    \caption{{\bf Positions of radio source PKS~2144$-$418  as measured in the ASKAP download that contained FRB~180924}. The positions are ASKAP (blue), the ATCA (purple), and VLBI positions from the Radio Fundamental Catalog \cite{Petrov19}. The radius of the circles are the 1$\sigma$ uncertainty for the ATCA and ASKAP positions - no systematic component has been included.Two DES positions of the optical counterpart are plotted. 
    The {\em measured} being the position given in the catalog, and the {\em corrected} is after the Gaia-DES offsets listed in Table \ref{tab:field_sources} have been applied. 
    For the DES positions the 1 sigma radius of the circle is the astrometric precision of 151 mas  \cite{DESDR1}.  }
    \label{fig:2144}
\end{figure}

Of the three ASKAP-detected sources, only PKS~2144$-$418 was detected in the ASKAP data with sufficient S/N to enable a useful measure of the astrometric uncertainties. The measured positions for this source from ATCA, ASKAP, VLBI and DES are shown in Figure \ref{fig:2144}. For this source, the offset between the ATCA and ASKAP positions is $+25.3 \pm 93.5$~mas in R.A. and $+47.2 \pm 91.5$~mas in Decl., where the 1$\sigma$ errors are the measurement errors for ASKAP and ATCA positions (dominated by the ASKAP positions), added in quadrature. We find no evidence for a systematic offset larger than the measurement errors when comparing the ASKAP and ATCA astrometry. Any systematic error is likely smaller than the measurement uncertainties quoted in Table 1. We therefore adopt a conservative estimate of systematic errors in ASKAP positions of 90~mas.

For FRB~180924,  we add this systematic error in quadrature to the 70 mas measurement error and obtain 120 mas uncertainties in the position of FRB~180924.

We also compared the ASKAP and ATCA positions of  PKS~2144$-$418 with a VLBI-measured position from the radio fundamental catalog (RFC version 2018c, \cite{Petrov19}). Asymmetric morphology can cause shifts in the radio positions measured with different angular resolutions, but in this case both the ATCA and ASKAP positions agree with the VLBI position within 50~mas, providing a further confirmation that our estimate of 90 mas for the systematic uncertainty is robust, and that our final position for the FRB is firmly anchored in the ICRF.

Finally, the position of the optical counterpart for PKS~2144$-$418,  from the DES catalog, is also shown in Figure  \ref{fig:2144}. While the uncertainty in the DES position is much larger than that for the radio sources, we obtain agreement between the optical and radio positions after applying the required shift to the DES frame (see below).

\subsubsection{Comparison between incoherent and interferometric localization methods}\label{sec:compare_fly}

\begin{figure}[htpb!]
    \centering
    \includegraphics[width=\linewidth]{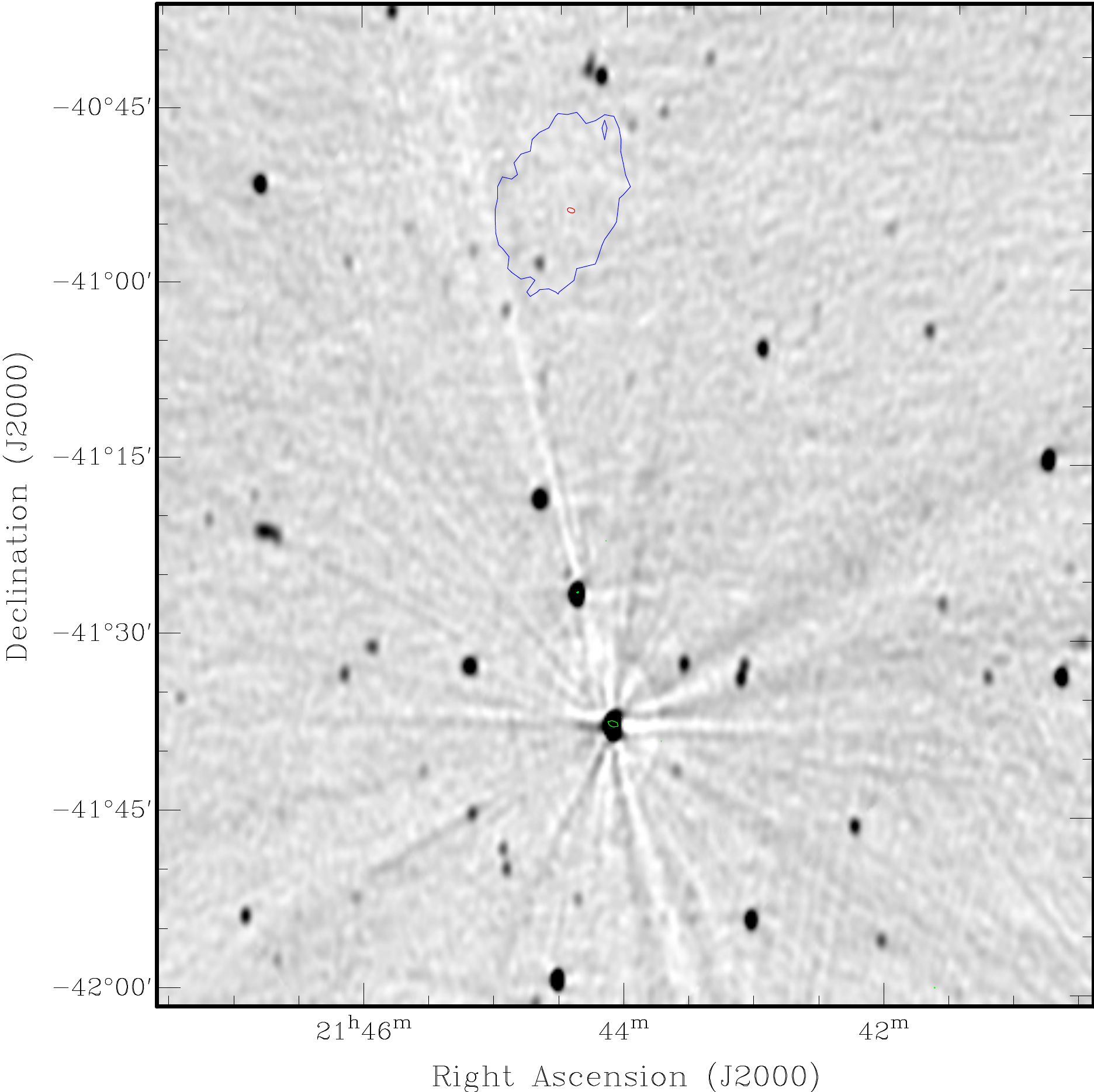}
    \caption{{\bf Comparison of incoherent and interferometric burst localization. } The blue contour shows the posterior localization region using the incoherent burst localization method previously employed \cite{Bannister17,Shannon18,Macquart18}.  The red circle is centered on the interferometric burst localization region.  
    The background is a radio image of the field taken from the Sydney University Molonglo Sky Survey (SUMSS) \cite{Bock99}. 
    The green circles are contours of radio-continuum sources in the ASKAP $3.1$\,s image.}
    \label{fig:incoherent}
\end{figure}

The interferometric detection enabled the test of a previous incoherent detection algorithm used to localize bursts. 
ASKAP FRBs are typically detected in multiple beams as the PAF beams oversample the focal plane.   By using the detected signal to noise ratios and the beam positions, it is possible to get a localization precision $\Delta \theta \approx \theta_{\rm FWHP}/{\rm S/N}\approx 10$~arcmin (90\%), where $\theta_{\rm FWHP}$ is the beam width at half power ($1^\circ$ for the observations here) and S/N is the burst signal to noise ratio in the strongest detection. 
This algorithm was employed on previous ASKAP detections and has been described in detail  \cite{Bannister17}.  
FRB~180924 was detected strongly (S/N$>$7) in two beams and had moderate detections (S/N$>$4) in two additional beams. In all beams adjacent to the primary detection the burst had a measured S/N$>$2 (The algorithm for measuring S/N is two sided. Beams distant from the burst position on the sky would have 50\% probability of being negative).
The burst position derived from this algorithm is consistent with the arcsecond localization (see Figure \ref{fig:incoherent}). The maximum likelihood position derived using the incoherent technique is only $30$~arcsec distant from the interferometric position, so is much smaller than the quoted $10$~arcmin uncertainty on the incoherent position.   This could be the result of good fortune, or of conservative assumptions applied in the incoherent algorithm.  In particular, the algorithm  incorporates uncertainties in beam sensitivities, shapes and positions that may be overestimated.

\subsection{FRB Polarimetry: Calibration}
Full-Stokes data products were produced by DiFX as described above. This provided datasets that had been calibrated in Stokes $I$.  However the bandpass calibration does not calibrate the other Stokes parameters.   The largest polarization defect in the observations is an unknown phase and delay between the two linearly polarized beams. The delay will cause a frequency dependent conversion of linear to circular polarization (in particular from stokes $U$ into $V$), and vice versa.   The measured values of U and V in the case of a phase offset $\Phi_0$ and a delay $\tau$ at a frequency $f$ are 

\begin{equation}
 U^{\prime}(f) = U(f) \cos \left( 2 \pi f \tau + \Phi_0 \right) + V(f) \sin \left(2 \pi f \tau  + \Phi_0\right)
   \end{equation}
 and
\begin{equation}
V^{\prime}(f) = -U(f) \sin\left(2 \pi f \tau + \Phi_0 \right)    + V(f) \cos\left( 2 \pi f \tau  + \Phi_0 \right)  
\end{equation}

We calibrated the data products by comparing observations of the Vela pulsar taken with ASKAP and those taken with Parkes. The pulsar has well defined polarization properties \cite{Johnston18}, with the pulse showing strong  ($\approx 95$\%) linear polarization with modest Faraday rotation (RM$=40$ rad m$^{-2}$), and approximately 5\% circular polarization.  

We formed a spectrum of the Vela pulsar from data taken with ASKAP 10~h after the FRB was detected:  this is the  {\em Vela} data set described above. 
We produced a model for the frequency dependence of the pulse profile in this band using publicly available observations from the 64-m Parkes radio telescope, which can largely cover the ASKAP band.
The Parkes data were calibrated  for differential gain and phase, as well as feed ellipticity, and followed techniques described previously \cite{Manchester13}. 
We produced models for the expected relative values for $Q$, $U$, and $V$, relative to $I$.  

The model for the polarization for the Vela pulsar is:
\begin{eqnarray}
&&\frac{L}{I}(f) = \frac{\sqrt{Q^2(f) + U^2(f)}}{I(f)} = 0.95\\
&&\frac{V}{I}(f) = 0.05 \\
&&\frac{Q}{I}(f) = L(f) \cos\left( 2 \rm{PA}(f)\right) \\
&&\frac{U}{I}(f) = L(f) \sin \left( 2 \rm{PA}(f)\right).\\
\end{eqnarray}

The polarization position angle is modeled to be
\begin{equation}
\rm{PA}(f) =  0.09\,{\rm RM}(f_{\rm GHz}^{-2} - 1.4^{-2}) - PA_0,
\end{equation}
and ${\rm PA}_0$ is the polarization position angle at a reference frequency of 1.4 GHz.  We fit this to the Parkes data to find RM = $40(1)$ rad\,m$^{-2}$  and ${\rm PA}_0=-0.35$ rad.   While  higher than the cataloged value, the rotation measure of the Vela pulsar is known to vary \cite{Hamilton77}.

While in principle we know the orientation of the ASKAP beams on the sky, we use the polarization model to solve for this (which causes the leakage of Stokes $Q$ into $U$ and vice versa).
Thus the measured values of the Stokes parameters ($Q^\prime$, $U^\prime$, and $V^\prime$) relative to the model are:

\begin{eqnarray}
&&Q^\prime(f)  = L(f) \cos\left(2 {\rm PA}(f) + \Psi_0\right) \\
&&U^{\prime}(f)   = L(f) \sin\left( 2 {\rm PA}(f) + \Psi_0) \cos( 2\pi f \tau + \Phi_0 \right) + V(f) \sin\left(2 \pi f \tau + \Phi_0\right)\\
&&V^\prime(f) = -L(f) \sin \left(2 \rm{PA(f)} + \Psi_0) \sin ( 2 \pi f \tau+ \Phi_0  \right) + V(f) \cos \left (2 \pi f \tau + \Phi_0 \right) 
\end{eqnarray}

In practice, we solve these equations by first solving for $\Psi_0$ using $Q^\prime(f)$.  
We then calculate $\Phi_0$ and $\tau$ and using $U^\prime(f)$. 
We then apply this calibration solution to the FRB data to determine the FRB rotation measure.  

Bayesian methodology is used to calculate the rotation measure for the FRB using the calibrated data.  We measured the rotation measure by fitting the fractional polarization in the calibrated $Q_i$ and  $U_i$ for each channel.

The model values for $Q$ and $U$ are then:
\begin{equation}
\hat{Q}_{m,i}=L_i \cos\left(2 {\rm PA}_{\rm FRB}(f_i)\right)
\end{equation}
and
\begin{equation}
\hat{U}_{m,i}=L_i \sin\left(2 {\rm PA}_{\rm FRB}(f_i)\right),
\end{equation}
where the position angle is 
\begin{equation}
{\rm PA}_{\rm FRB}(f_i) = 0.09\,{\rm RM}_{\rm FRB} f^{-2}_{i, GHz} + {\rm PA}_0
\end{equation}

We assume the noise is identical in $Q$ and $U$, and is Gaussian distributed with no frequency dependence.

 The likelihood in this case is:
\begin{equation}
\mathcal{L} = \prod_{i}^{N_c} \frac{1}{2 \pi \sigma^2} \exp \left[- \frac{(Q_i-\hat{Q}_{m_i})^2}{2 \sigma^2} \right]    \exp \left[-\frac{(U_i-\hat{U}_{m_i})^2}{2 \sigma^2} \right] 
\end{equation}

We calculate the posterior probabilities for the parameters (${\rm PA}_0$, ${\rm RM}_{\rm FRB}$, and $\sigma$).  $L_i$ is a nuisance parameter so we analytically marginalize over it, assuming uniform priors.  
The final likelihood is then
\begin{equation}
\mathcal{L} = \prod_i^{N_c}  \frac{1}{2 \pi \sigma^2} \exp \left[- \frac{Q_i^2 + U_i^2 -\left( Q_i \cos(2{\rm PA}_{\rm FRB}) + U_i \sin (2 {\rm PA}_{\rm FRB})  \right)^2  }{2 \sigma^2} \right].
\end{equation}

We assume uniform priors on ${\rm PA}_0$, ${\rm RM}_{\rm FRB}$, and  Gaussian priors on  $\tau$  and $\Phi_0$, with the means of $\tau$ and  $\Phi_0$  set to their maximum-likelihood values from the Vela-pulsar fit and the standard deviations set to the fit uncertainties. We sample the posterior distribution using a nested-sampling algorithm \cite{Feroz09}.   
The method was tested by measuring the rotation measure of well known pulsars using observations from the $64$-m Parkes radio telescope.  
The frequency-averaged Stokes parameters (after correcting for Faraday rotation) can be found in Table \ref{tab:stokes_parameters}.

\subsection{Optical Astrometry}

We referenced positions of the host dark energy survey (DES, \cite{DESDR1})  galaxy to the optical reference frame using Gaia \cite{GaiaDR2}.    The Gaia reference frame is known to be well aligned with ICRF3 used for radio astrometry \cite{GaiaDR2}.
We use the DES image and catalog to conduct astrometry, rather than the deeper VLT image (discussed below), because the imaging camera used for DES has been well characterized \cite{Bernstein2017}.  The wide field of view ensures that many stars are within a single image, enabling a robust comparison of the DES astrometry from that of Gaia.

From the DES Data Release 1 (DR1) catalog, we selected sources in a 0.2 deg square around the FRB position that had a g-band magnitude less than 21 and were stellar or quasi stellar as determined using the cataloged SPREAD parameter. We selected sources as being quasi stellar using SPREAD $<$ 0.003.   In total, 111 sources in the DES catalog satisfied these criteria. We cross-matched the DES sources with the Gaia catalog and found unique identifications for all of them.   

The DES DR1 comprises data from the first 3 seasons of DES (starting in 2013.63 and ending at 2016.14, Ref. \cite{DESDR1}).  To account for proper motion, we referenced the Gaia sources (which have measured proper motions and positions defined at epoch 2015.5) to  DES, which we assumed to be observed at epoch 2014.9, which is the mean of the start of the first season and the end of the last season.  We found that accounting for source proper motion did not substantially affect the astrometric offset between DES and Gaia positions, but did reduce the root-mean-square scatter in the positions (Table \ref{tab:gaia_des_compare}). 
  
\begin{figure}[htpb!]
    \centering
    \includegraphics[width=\linewidth]{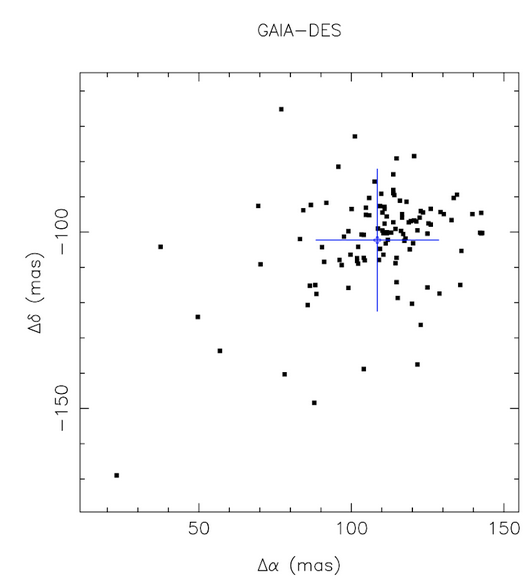}
    \caption{{\bf Apparent offsets in R.A. ($\Delta \alpha$) and Declination ($\Delta \delta$) from Gaia-DR2 and DES-DR1 catalogs.}  The black squares show the positions of individual sources.  The center of the blue cross is the mean of the positions and the size is the 1-sigma root-mean-square .}
    \label{fig:DES_Gaia_offsets}
\end{figure}

\begin{figure}[htpb!]
    \centering
    \includegraphics[width=\linewidth]{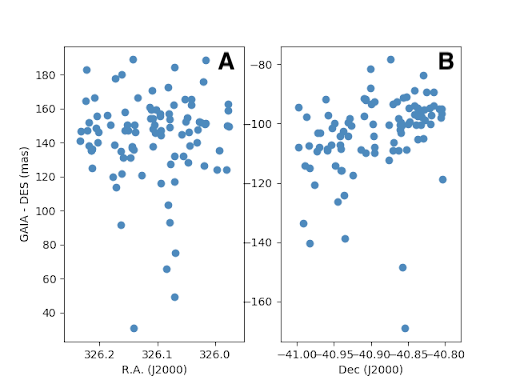}
    \caption{{\bf Apparent offset in positions from Gaia-DR2 and DES-DR1 catalogs.} (A) RA offset vs RA. (B) Dec offset vs Dec. }
    \label{fig:DES_OFFSETS_RADEC}
\end{figure}

We plot the R.A. offset and Decl. offset in Figure  \ref{fig:DES_Gaia_offsets}
 and find no evidence for any correlation between the offsets in axes. Similarly, plotting R.A. offset vs R.A. and Decl. offset vs Decl.  in Figure \ref{fig:DES_OFFSETS_RADEC} we find no evidence for large systematic errors in the position, other than a shift, between DES and Gaia positions in the 0.2 degree field we investigated.

The derived offsets and scatter in the positions are displayed in Table \ref{tab:field_sources} and applied to the optical positions of the host galaxy in Fig. 2 and the host galaxy of the VLBI check source in Fig. \ref{fig:2144}.

\subsection{FORS2 Optical imaging}
Observations of the host galaxy of FRB~180924 were carried out
in service mode  (i.e., queued observing) with the FORS2 instrument\cite{FORS2} mounted on Unit Telescope 1
(Antu) of the European Southern Observatory's Very Large Telescope
(VLT) on 2018~Nov~9~UT. Five exposures of 500~s in the g\_HIGH filter
with incremental offsets of 10$^{\prime\prime}$ between each
were followed by $5 \times 90$~s exposures in the I\_BESS filter
with similar offsets. The Standard Resolution collimator was
used with $2\times2$ binning of the CCD array to yield a pixel
scale of 0.25$^{\prime\prime}$ per pixel. Skies were photometric,
and the seeing varied between 0.7 and 1.3$^{\prime\prime}$.

Standard image processing was performed using version 2.9.1 of
the ESO Reflex\cite{ESOReflex} pipeline.
The dithered images in each filter were registered and co-added
using a custom script and tasks in v2.16 of {\sc  iraf}. The
zero-point for $I$-band was taken from ESO's nightly monitoring
program, while DES $g^{\prime}$-band imaging of the surroundings
to FRB~180924 was used to calibrate the g\_HIGH imaging. Summing up the
flux from the FRB host galaxy, and using transformations between
the $V+I$ and $g^{\prime}+i^{\prime}$ filters\cite{Blanton2007}
we find
$i^{\prime} = 20.28 \pm 0.05$ and $g^{\prime} = 21.57 \pm 0.04$.
For comparison, the integrated magnitudes from the DES Catalog \cite{DESDR1}
are $i^{\prime} = 20.14 \pm 0.02$ and $g^{\prime} = 21.62 \pm 0.02$.

Astrometry for the FORS2 mosaics in g- and I-band was refined using astrometry.net \cite{Lang2010}  and the United States Naval Observatory (USNO) B1.0 catalog \cite{Monet03}.

\subsection{Keck-KCWI Integral-field spectroscopy}

On UT 04 October 2018, we executed a Target of Opportunity observation of the field surrounding FRB~180924 with the Keck Cosmic Web Imager (KCWI; Ref. \cite{Morrissey18})  instrument on the W.M. Keck II telescope. This integral field unit (IFU) spectrometer was configured with its medium slicer giving an approximately $16$ arcsec $\times$ $20$ arcsec field-of-view and the instrument was rotated to $10$ deg. west of the parallactic angle.  
We employed the BL grating tilted to a central wavelength of approximately $4500$ Angstrom given a resolving power of ~1,800 and wavelength coverage spanning $\sim 3500-5500$ Angstrom.   Beginning at 06:47 UTC, we obtained a series of four, $600$-s exposures with small ($\sim$ 1 arcsec) offset dithers in-between.  
The data were taken under good conditions but at a high airmass ($\sim  2.0$). 
A series of calibration images were taken during the preceding afternoon.

The spectral images were processed with the KCWI data reduction pipeline v1.1.0 \cite{kderp} which solves for the wavelength and geometric solutions to generate a sky-subtracted IFU data cube.  Using a sensitivity function derived from an observation of the flux standard star BD+284211, the data cubes were then flux calibrated.  Each spaxel in this cube has a rectangular dimension of approximately 0.7 arcsec $\times$ 0.29 arcsec.  Using the Montage software package \cite{Berriman17}, we then combined the image frames, and produced a final cube with square 0.29 arcsec $\times$ 0.29 arcsec pixels aligned with the parallactic. Lastly, we extracted the galaxy spectrum using the 5 $\times$ 5 spaxels surrounding its center, summed without weighting (i.e. boxcar).  The wavelengths of this spectrum were corrected to vacuum and into the heliocentric reference frame.  From the observed [O~\textsc{ii}] emission and Ca H+K absorption, we estimate the galaxy redshift to be $z$ = 0.3214 and estimate an uncertainty of approximately 30 km\,s$^{-1}$ due to asymmetries in the spectral features and systematics in the wavelength calibration.

\subsection{Gemini-GMOS long-slit spectroscopy}

On 2018 Oct 5 UT we triggered Target of Opportunity (ToO)
observations of two galaxies (B and C in Figure \ref{fig:host2} of the main text) located near the localization
of FRB~180924 with the
GMOS spectrograph \cite{GMOS-N, GMOS-S} mounted on the Gemini South telescope. 
Specifically, we centered a long slit (1\,arcsec width)
on galaxy~A, oriented with a position angle of PA=$280.9$\,degrees
to cover the neighboring galaxy~B.
Observations were carried out in dark time, airmass below 1.08, seeing between $0.7-1.0$\,arcsec,
and thin cirrus.
Four individual exposures of $700$\,s were taken
using the R400 grating centered at $7000$\,\AA\ with the blocking filter GG455 and a $2 \times 2$ CCD binning. 
This configuration gives a spectral coverage of 
$4750-9300$\,\AA\ at a resolving power of $R\equiv \frac{\lambda}{\Delta \lambda}\approx 1000$ at $\lambda=7000$\,\AA, respectively, and with a dispersion of $\approx 1.5$\,\AA\ per pixel. 
Calibration images for flat fielding, bias subtraction, and wavelength calibration were also acquired. These spectral images were reduced using the PypeIt package \cite{pypeit}
using standard techniques for image processing, extraction,
calibration, and coadding.  

From the GMOS data (Figure \ref{fig:host2}C-D ) we confirmed
the redshift of each galaxy as first estimated from the
Keck observations.  These data also 
demonstrated the relatively weak H$\beta$ emission
for galaxy~A (suggesting internal extinction)
and an unusually high [NII]/H$\alpha$ flux ratio (suggesting an AGN or LINER emission). 

\subsubsection{VLT-MUSE integral-field spectroscopy}
We obtained data from the Multi Unit Spectroscopic Explorer (MUSE)\cite{MUSE} mounted on UT4 (Yepun) of the VLT
on 2018 Nov 5~UT, in gray time, airmasses between $1.14-1.22$ and DIMM seeing between $1.1-1.3$\,arcsec. Four individual exposures of 628~s were taken in the MUSE Wide Field Mode - Adaptive Optics (WFM-AO) mode, roughly centered on galaxy~A. Each exposure was taken with a small spatial dithering offset of 1\,arcsec and at a position angle (PA) offset of $+$90 degrees with respect to the previous one starting with a PA~$=0$\,deg, to reduce the effects of bad pixels and difference in illumination between the individual image slicers in a given spaxel. 

The individual exposures were reduced and combined with the standard European Southern Observatory (ESO) MUSE pipeline version 2.4 \cite{Weilbacher14}, to produce a combined datacube of total exposure time of 2512~s. The wavelength coverage of the datacube is $4750-9300$\,\AA\ at a resolving power of $R\equiv \frac{\lambda}{\Delta \lambda}= 1750-3590$, respectively, and with a dispersion of $1.25$\,\AA\ per pixel. The spectral range between $5800-5950$\,\AA\ has been blocked out to avoid contamination from the NaD laser emission used in the AO mode. The final field of view is about $1\times1$\,arcmin$^2$ with a spatial pixel scale of $0.2$\,arcsec per pixel. Given the WFM-AO mode, the effective seeing of the combined datacube is $\approx 0.8$\,arcsec. In order to improve the sky subtraction we run the Zurich Atmospheric Purge (ZAP) pipeline \cite{ZAP} version 2.0, masking out bright objects before defining the overall sky level. 

The MUSE astrometric solution from the standard pipeline showed a systematic offset of about $\approx 2$\,arcsec, which we corrected as follows. First, we created a mock $r$-band image from the MUSE datacube, and then imposed the astrometric centroid of galaxy~A from this image to match that of the DES $r$-band image. In this manner, we made the MUSE astrometry to match that of DES. Then, we applied an extra smaller offset to pass from the DES astrometry to that of Gaia-based \cite{GaiaDR2} astrometry described above ($+108$\,mas in R.A. and $-97$\,mas in Declination).

The MUSE data allowed for a more complete spectroscopic search for sources at the position of FRB~180924. We confirmed the redshifts of galaxies~A and B, as well as a third source about $4.5$\,arcsec to the North East of galaxy~A, referred to as galaxy~C, having $z=0.50055$ (see Figure~\ref{fig:host2}). A fourth source about $10$\,arcsec to the West of galaxy~A has a larger redshift of $z=0.75$ and is not considered further. In summary, MUSE observations confirmed that galaxies~A, B and C are the only sources within a radius of $5$\,arcsec from the FRB position to a conservative limit of $r\sim 25$, also corroborated by the deep $g^{\prime}$ deep imaging (Figure~\ref{fig:host2}).

\subsection{Searches for repeated pulses}

Observations of the field were taken with the 64-meter Parkes radio telescope, on 2018-10-02 for $8.8$~hr and on 2018-10-16 for $2$~hr.   The telescope was pointed at the position derived using the fly’s-eye localization method presented previously \cite{Bannister17} as the sub-arcsecond position had not been determined at the time of the observations.  As discussed in Section~\ref{sec:compare_fly}, this position (and hence the telescope pointing) was offset from the true FRB position by approximately 30 arcsec, much smaller than the nominal $10$~arcmin uncertainty quoted for the fly's-eye method, and much smaller than the Parkes beam.  

Observations were conducted with the 20-cm multibeam receiving system \cite{Staveleysmith96}.   Data were recorded with the Berkeley-Parkes Swinburne Recorder  \cite{Keith2010}, with  $64~\mu$s time resolution and $380$~kHz spectral resolution, over a $350$~MHz use-able bandwidth The data were searched in real-time for radio pulses using {\tt Heimdall} \cite{Barsdell12}.   No pulses were detected above a S/N of $10$ at the dispersion measure of the burst nor at any other dispersion measure between 10 and $4000$ \pccm. The system  system equivalent flux density of the central beam of the multibeam system is approximately $40$~Jy, so the limiting fluence for a pulse of with $w_{\rm ms}$ is $0.5 w_{\rm ms}$~Jy\,ms.

\subsection{ATCA 4-cm observations}

Observations of the field were taken with the ATCA at 5.5 and 7.5 GHz to search for compact, continuum emission associated with the FRB. Initial observations were carried out 24 hr after the detection of the FRB. Since these observations were taken prior the sub-arcsec determination of the burst position, three pointings were observed to cover the localization region determined by the incoherent array detection method applied to previous fly's-eye detections with ASKAP \cite{Bannister17,Shannon18}.    The array was in the 750C configuration (5 of the 6 antennas within $750$-m of one another, with the remaining antenna approximately $3$-km distant) leading to a resolution of  of 7.8 $\times$ 1.1 arcsec at $5.5$\,GHz and 5.7 $\times$ 0.8 arcsec at $7.5$\,GHz using Briggs weighting with robust$=0.5$. The total observation duration (including calibration and overhead) was $8.5$\,hr.  No continuum source was found at the location of the FRB or anywhere in the host galaxy. 

Follow-up observations in the more extended 6A configuration (maximum baseline length of 6 km) were carried out at the same frequencies on 2018 October 4, 10 days post-burst. 
A single pointing was observed at the known position of the FRB for 5 hrs reaching rms of $11.8$, $10.3$ $\mu$Jy at $5.5$ and $7.5$ GHz. The radio galaxy, PKS~1934$-$638 was used as the bandpass and primary calibrator for both observations, while PKS~2106$-$413 and PKS~2211$-$388 were used for phase calibration for each observation respectively. 
Again, no continuum source was detected at the location of the FRB or in the host galaxy.

The calibrated visibilities from these two observations were combined and re-imaged using natural weighting to achieve a spatial resolution of 8.0$\times$2.2 arcsec and an rms of 6.8 $\mu$Jy\,beam$^{-1}$ at the position of the FRB. 
Using a redshift of $z = 0.32$ and assuming a flat spectral index this results in a $3\sigma$ upper limit on the radio spectral luminosity of $L_{\nu} < 5.5 \times 10^{21}$~W\,Hz$^{-1}$ at 6.5~GHz, a factor of $3.5$ times smaller than luminosity of the persistent source associated with FRB~121102 \cite{Chatterjee17}.

\subsection{ASKAP radio-continuum observations}

We observed the field in the vicinity of FRB~180924 during commissioning of ASKAP on 2018 September 26 10:01 (UTC). The duration of the observation was 10\,h with a total of 13 ASKAP antennas at a central frequency of 1344\,MHz and a bandwidth of 288\,MHz.  As FRB~180924 was situated within 12~arcmin of the center of beam 18 of the ASKAP \emph{closepack36} footprint, only that beam was processed. 
The bandpass and flux scale was calibrated using observations of the primary calibrator PKS B1934$-$638 with  {\sc CASA} \cite{CASA}. 
Deconvolution was performed with the {\sc CASA} task {\tt clean}  using the Briggs weighting scheme (robust$=0$). The resulting image was restored with a $35$~arcsec $\times$ $30$~arcsec beam (position angle 34$^\circ$ ) and achieved a $1\sigma$ root-mean-square sensitivity of 150\,$\mu$Jy beam$^{-1}$.

\subsection{Host-galaxy analysis}

\subsubsection{Morphology}

We have assessed the stellar morphology of the host galaxy, using the VLT/FORS2 data,  with the
{\sc GALFIT} software package \cite{galfitv3}. We used the image of a
nearby star as an estimate of the point spread function (PSF).
It has a DES $r$-band magnitude of $18.656 \pm 0.002$ and was chosen because
it was the closest star that was not saturated in the DES imaging. The
same star was also used for analysis of our MUSE and FORS2 data. We fitted the galaxy with a S\'ersic profile and the sky background with a constant value. The
sky background fit was $\sim 0.1\%$ of the peak value of the galaxy
surface brightness. Analysis of the DES $r$-band image yields an effective
radius $r_{\rm eff} = \reffv$\,kpc which is within the uncertainty of the Nyquist sampling (2 pixels),
i.e.,\ the galaxy is not well-resolved.
The analysis further yields
an estimated S\'ersic index $n = 1-2.4$
with $n \approx 2$ preferred.  These
span the range from an exponential disk to
a more bulge-dominated system.
We also measure an axis-ratio $(b/a) = 0.62 \pm 0.05$
indicative of a disk-like profile and a position angle of
PA~$= -24 \pm 5$\,deg.
Adopting this model, the FRB event occurred at 
a radius that encompasses $\approx 90\%$ of the stellar
light.

We have repeated the {\sc GALFIT} analysis with a pseudo-narrow band (NB) image
at H$\alpha$ generated from the MUSE data cube and recover 
similar results but with a lower central S\'ersic index ($n=1.7$
is preferred). Based on the {\sc GALFIT} residuals, we can also conclude that the the galaxy does not show any obvious excess or absence of H$\alpha$ flux at the position of the FRB~180924 with respect to the overall smooth distribution modeled. 

The same exercise was also performed for the FORS2 I band data and this gave us tighter bounds on the index, $2.0\pm0.2$. The estimated light fraction within the radius corresponding to the FRB is $0.92\pm0.02$.
We caution again, that the galaxy is not well-resolved by these ground-based images with typical seeing of $0.8-1.3$\,arcsec.

\subsubsection{Photometry}

The host galaxy of FRB~180924 is detected in all five bands of the
DES DR1 survey \cite{DESDR1}.  The galaxy is also detected in the W1 and W2 bands
of the WISE all-sky survey.  Table~\ref{tab:photom}\ summarizes this
set of photometric measurements, all of which were taken from
the DES DR1 release and its cross-matched WISE catalog \cite{Wright10}. 
The galaxy has a $g-i \sim 1.5$ color (roughly
corresponding to $u-r$ rest-frame at $z=0.32$), which is indicative
of a galaxy with limited recent star formation.

We performed an SED analysis of the galaxy photometry with the
CIGALE software package \cite{Noll:2009aa}.  The grid of SEDs that
defined the model search space includes allowances for an
AGN contribution \cite{Dale:2014aa} and nebular emission.  The stellar population models \cite{Bruzual:2003aa}, use a `periodic'
star-formation history (SFH) model characterized by a number of exponentially decaying starburst episodes assuming a standard \cite{Chabrier:2003pd} initial mass function (IMF).
Both the stars and gas are attenuated by model composted of standard extinction curves (\cite{Calzetti:2000lr},\cite{Leitherer:2002ab}) with a 2175 \AA\ bump feature.

Figure~\ref{fig:host} compares the best-fitting model from CIGALE
with the observed fluxes, demonstrating agreement
with the data.  
This model is dominated by 
stellar emission from the population of older stars, both
by mass and luminosity (at optical and near-IR wavelengths).
The SFH-weighted age is 5.5\,Gyr and the model Star Formation Rate (SFR)
exhibits a monotonic, decreasing rate over the past 10\,Gyr.

The stellar mass is estimated at $\stellarmass$ for the
assumed IMF.
The fitting results in a non-negligible selective extinction, 
$E(B-V) \approx 0.3$\,mag, for the dominant, old stellar component.
Table~\ref{tab:cigale}\ summarizes the best-fitting parameters and
estimated uncertainties from the Bayesian methods used
in CIGALE.  There are potentially
substantial degeneracies in the results, e.g.,\ between
dust extinction and estimated star-formation rate.
Furthermore, the results have modest dependence on 
the allowed
parameterization of the SFH.   For example,
a model which combines an old-and-dead stellar component with a
younger, recently formed component yields an 
$\approx 30\%$ lower stellar mass.
Nevertheless, the preferred model is an 
old galaxy with a declining SFH and low 
current SFR (consistent with zero; see below for 
an upper limit derived from H$\alpha$ line-emission). 
The best-fitting model includes a small AGN component, but
its estimated contribution is consistent with zero.

\subsubsection{Spectroscopy}

Our spectroscopy includes long-slit observations with
the Gemini-S/GMOS spectrograph and IFU datacubes with
the Keck/KCWI and VLT/MUSE.
Of these, the VLT/MUSE dataset has the highest S/N over
the widest wavelength coverage and we focus on that dataset; results are consistent between spectrometers.
From the MUSE datacube, we constructed a 1D spectrum of
the host galaxy by optimally coadding its spaxels 
weighting by the total flux in the white-light image and renormalizing to conserve total flux within an aperture. For this we used the PyMUSE software \cite{pymuse}, defining an elliptical aperture 
 comprising the full spatial extent of the galaxy~A but avoiding flux from galaxies~B and C. This optimal extraction is robust to changes in the edge limits of the aperture, because spaxels with total flux consistent with the sky level have effectively zero weight.
We then employed the pPXf software package \cite{Cappellari:2004aa,Cappellari:2017aa}
 on the extracted 1D spectrum.
The software fits a set of stellar population models
and nebular emission lines to an input spectrum to
assess the kinematics and stellar composition of the galaxy.
For the stellar spectra, which vary in age and metallicity,
we employ models from a database 
\cite{Vazdekis:2010aa} that assume the Chabrier initial mass function.
The model also allowed for a 3rd order, additive polynomial
to adjust the continuum to account for minor fluxing errors
and extinction of the stellar light by a foreground
screen of dust.

Consistent with the CIGALE analysis describe above,
the best-fitting pPXf model is dominated by old stars
with a light-weighted age of $t_{\rm weighted} \approx 6$\,Gyr.
The model also results in a selective extinction 
$E(B-V) \approx 0.2$\,mag (from the stellar continuum),
and the light-weighted 
mass-to-light ratio implies a total stellar mass of 
$\approx 10^{10} M_\odot$.

The pPXf package also estimates line fluxes for a series
of standard nebular emission lines after correcting for
any coincident stellar absorption.
These are listed in Table~\ref{tab:tblines} 
with their estimated uncertainties.
From the observed Balmer decrement, 
$\rm{F(H\alpha)/F(H\beta)} = 3.8 \pm 0.1$, where $F()$ is the flux of the line,
we infer a modest internal extinction (the Galactic
extinction along this sightline is estimated to be
less than 0.1\,mag \cite{Cardelli:1989lr}).
Adopting a standard Galactic extinction law
\cite{Cardelli:1989lr} and the Case~B recombination intrinsic ratio
of $\rm{H\alpha/H\beta} = 2.8$ \cite{Osterbrock:2006fk},
we estimate $A_V = 1.0$ and $E(B-V) = 0.3$.
The extinction-corrected line-luminosities are listed
in Table~\ref{tab:lines}\ using our adopted cosmology.

From these line luminosities, we estimate additional
properties of the host galaxy.  First, we note the
[NII]/H$\alpha$ line-ratio exceeds the values observed
for normal, star-forming galaxies.  This is emphasized
in Figure~\ref{fig:BPT}\ which presents a canonical Baldwin-Phillips-Terlevich
(BPT) diagram frequently used to assess the 
ionizing sources within a galaxy.  We find that
the host galaxy of FRB~180924 is offset
from the locus of $z \sim 0$, star-forming galaxies.
Instead, its position in Figure 4 is near the locus of galaxies
known to host active galactic nuclei (AGN) and those
that exhibit LINER emission \cite{fernandes10}.

We conclude that the line emission in this
galaxy is produced in part by the ionizing mechanisms
invoked for LINER galaxies.  For galaxies dominated by
old stellar populations, the currently favored scenario
is starlight from post-AGB stars.  Such stars 
might also provide the dust responsible for the observed
extinction.

If the observed H$\alpha$ flux were dominated
by the line-emission of star-forming regions, the
luminosity would yield an estimate of the
current SFR. 
Applying the standard conversion factor \cite{Kennicutt2012},
this would imply
$\approx 2 M_\odot \, {\rm yr^{-1}}$ with uncertainty
dominated by the applied extinction correction 
($\approx 30\%$).   
Given the LINER excitation, the derived value
yields only an upper limit for the galaxy.  
We therefore report an upper
limit of SFR~$< 2 M_\odot \, {\rm yr^{-1}}$.

\section{Supplementary Text}

Throughout the analysis we use cosmological parameters from the Planck 2015 results \cite{2016A&A...594A..13P}.

\subsection{Arguments against an association with the background galaxies}

Given the association of FRB\,121102 with a dwarf galaxy, it is necessary to consider whether the burst source can be associated instead with two other galaxies, which are in the background to DES~J214425.25$-$5400.81. These galaxies are labeled label B ($z=0.384)$ and C ($z=0.50005$) in Figure \ref{fig:host2}.  DES~J214425.25$-$5400.81 is labeled A.     
Two separate arguments suggest that FRB\,180924 does not reside in the halo of either of these galaxies.

Firstly, if we assume that the burst originated in the halo of galaxy B or C, then the probability of a chance coincidence  is at most 7\% for galaxy B and  4\% for galaxy C under the unlikely assumption that the burst could originate anywhere within the halo of the background galaxy with uniform probability to the observed separation from the galaxy center.  If the probability was (for instance) weighted by the stellar light of galaxy B or C, the likelihood of chance coincidence becomes negligibly small for either galaxy.
Secondly, the burst properties are inconsistent with it having propagated from a background galaxy through the foreground.    
Integral field spectroscopy shows that ionized gas is present in the foreground galaxy at the position of FRB\,180924.  
Any burst emitted from either other galaxy would incur  sufficient  excess dispersion
from its passage through the ionized gas in the lower-redshift galaxy to make it inconsistent with that expected from the intergalactic medium.
In this scenario, the burst would also be expected to be severely broadened due to diffractive scattering,  if the gas in the lower-redshift galaxy has turbulence comparable to that in the Milky Way. 
We therefore conclude that the burst is associated with the closest galaxy, DES~J214424.97$-$405400.2 (galaxy A).   

\subsubsection{The dispersion measure contribution of galaxy A}

At an offset of $\approx 3.9$\,kpc from the center of 
a relatively massive galaxy, the dispersion measure contributed
by the host galaxy of FRB~180924 and its halo are likely to
be substantial.  Consider first a fiducial estimate for gas within
the dark matter halo surrounding the galaxy.
Adopting a stellar mass of $2 \times 10^{10} M_\odot$,
we may roughly estimate the halo mass using the halo
abundance matching technique \cite{moster10}.
This methodology yields a halo mass $\log_{10} M_h = 11.9$
with an $\approx 0.3$\,dex uncertainty.
In the following, we also assume a concentration 
$c=7.7$ which is characteristic of halos of this mass
at low redshift.   We make the following assumptions
on the baryons within the halo:
(i) the density profile of halo gas follows a modified Navarro-Frenk-White (NFW) profile:
\begin{equation}
\rho_b = \frac{\rho_b^0}{y^{1-\alpha} (y_0 + y)^{2 + \alpha}} \;\; ,
\end{equation}
as discussed in \cite{xyz19} where $\rho_b^0$ is a normalizing density, $y$ is dimensionless radial coordinate, with a steepness index set to $\alpha=2$ and (dimensionless) scale length $y_0=2.$”;
(ii) the halo contains a cosmic fraction of baryons, 
i.e.\ $M_b = M_h (\Omega_b/\Omega_m) (f_b/1)$ where $M_b$ is the baryon mass, $M_h$ is the halo mass, $\Omega_b$ is the cosmic average baryon density, $\Omega_m$ is the cosmic average matter density, and $f_b$ is the fraction of baryons in halos,
and that 75\%\ of these baryons are in a diffuse,
ionized plasma within the halo.  This sets the
normalization in the above equation.
We emphasize that neither $f_b$ nor the gas-density profile
are well determined for galaxy halos \cite{Ioka2003,Inoue04,McQuinn14,Deng2014,xyz19}.

Adopting this fiducial model and an impact parameter $R_\perp = 3.9$\,kpc, the integrated electron column density through the halo defined by 
its virial radius ($r_{\rm vir} \approx 190$\,kpc)
is $N_e = 108$\pccm.  This gives a halo-integrated dispersion
measure ${\rm DM}_{\rm A}^{\rm halo} = 82$ \pccm.
Of course, if FRB~180924 is hosted in the stellar disk of galaxy A it will be near the mid plane of the halo and the expected halo contribution would be half of this quantity.
We caution that this estimate is subject to large uncertainties
related to both $f_b$ and the adopted density profile.  
Numerical simulations of galactic halo gas predict $f_b$
values that range from unity to much less than one, often
with a steep halo mass dependence \cite{fielding+17,Hafenetal2018}.
This implies at least a 50\%\ uncertainty in our adopted 
${\rm DM}_{\rm A}^{\rm halo}$ value.
This treatment ignores any gas local to the
burst source.  Given the low flux of H$\alpha$ emission at the
FRB location, however, we infer this to be a small contribution.

\subsubsection{The dispersion-measure contributions to galaxies B and C}

We consider whether the observed DM of 362\pccm\  of FRB\,180924 is sufficiently large to be compatible with passage through all of the Milky Way (estimated as ${\rm DM}_{\rm MW} \gtrsim 70\,$\pccm\, as described above, the halo of galaxy A (${\rm DM}_{\rm A} \approx 82\,$\pccm), and through the intergalactic medium out to the redshifts of galaxies B or C, at $z=0.384$ and $z=0.50055$ respectively.  We use models \cite{McQuinn14} to quantify the minimum likely contribution from the IGM. These models account for variation in the IGM dispersion measure caused by variations in the distribution of cosmic baryons between different sight lines, for which much of the variance is due to the number of galaxy baryonic halos intersected. 
We use as a representative model, one in which each galaxy halo baryon profile follows an NFW distribution that extends out to the virial radius of each galaxy.  The NFW halo models here are conservative because most models find the profile is more diffuse than NFW (where NFW traces the dark matter), which would lead to the distribution being narrower.

\begin{figure}[htbp!]
    \centering
    \includegraphics[width=\linewidth]{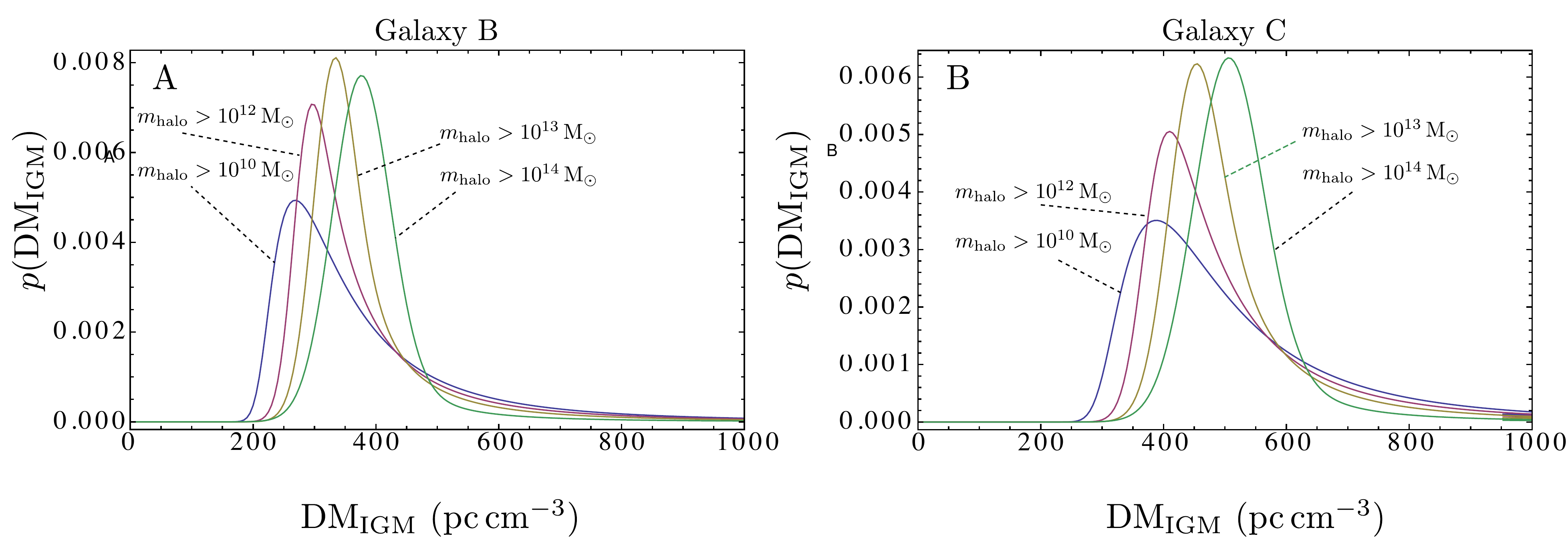}
    \caption{{\bf The probability distribution of the IGM contribution to the dispersion measure, $p({\rm DM}_{\rm IGM})$}. (A) shows the distribution if the burst emanates from either galaxy B  and (B) shows the distribution comes from Galaxy C, under the assumption that the baryonic halos of galaxies near the line of sight follow an NFW profile extending out to the virial radius.  Differences in the probability distributions reflect uncertainty in the minimum halo mass.    
    } \label{fig:IGMhalos}
\end{figure}

Figure \ref{fig:IGMhalos} shows the expected DM distribution parameterized according to the minimum halo mass for which a galaxy will retain its baryonic halo; we consider halo masses in the range $10^{10}$-$10^{14}\,$M$_\odot$.  The lowest minimum DM is obtained for models in which the minimum halo mass is $10^{10}\,$M$_\odot$.  The physical motivation for parameterizing the models in terms of a mass cutoff at some minimum mass is that below this minimum mass feedback is able to blow out all the halo gas; this is thought to occur around 
$10^{12}\,{\rm M}_\odot$ \cite{Hafenetal2018}, but this does depend on the specific feedback implementation.   

A useful constraint is set by considering the location of the fifth percentile of the DM distribution (i.e. the point above which 95\% of the probability is contained).  For $z=0.38$ this minimum IGM DM contribution is 227\pccm, while for $z=0.50$ it is 325\pccm.  Folding in additional contributions from the Milky Way and galaxy A, the minimum expected DM from a burst associated with 
galaxy B is 380\pccm, while it is 480\pccm\ 
for one associated with galaxy C.  Both lower limit estimates are inconsistent with the measured burst DM.

\subsection{Stars and their host galaxies}
The stellar mass function for nearby galaxies \cite{baldry2012} implies that most stars in the nearby Universe lie in moderately massive galaxies with stellar masses above about 10$^{10}$ M$_{\odot}$. Table \ref{tab:stars} shows a rough breakdown of the fraction of stars in the local Universe that lie in galaxies of different masses - the distribution is strongly influenced by the 'knee' of the galaxy stellar mass function \cite{baldry2012}, which lies near $4\times10^{10}$ M$_\odot$. Above this mass the volume density of galaxies starts to decrease rapidly, while below this mass the volume density of galaxies rises too slowly to offset the smaller number of stars in each galaxy. 
Unlike FRB 121102, FRB 180924 is located in the kind of moderately massive galaxy in which the bulk of stars in the local Universe are found. If this is a common property of FRB host galaxies, it would suggest that the progenitors of FRBs are drawn from the overall stellar population (including the endpoints of stellar evolution, like neutron stars and white dwarfs), rather than from an exotic sub-population found mainly in dwarf galaxies.

\subsection{Models of burst dispersion measure:  IGM and host residual}

Models of the baryon distribution of the IGM generically show there is a minimum non-zero dispersion measure contribution from the IGM for any object at $z > 0.2$. 
This can be used to place an upper limit on the host galaxy dispersion measure contribution in the present case.  We derived the probability distribution of ${\rm DM}_{\rm IGM}$ for a variety of IGM models previously considered \cite{McQuinn14}. 

Specifically, we examined models in which the baryonic halos of galaxies that intersect the FRB sightline have gas profiles whose densities possess an NFW distribution \cite{Navarro96} around their galaxy centers. 
The baryonic profiles were taken to extend out to the virial radius and were parameterized in terms of the minimum halo mass that retain its gas: we considered minimum halo masses over the range $10^{10}$ to $10^{14}$ solar masses, ranging between dwarf galaxies and galaxy cluster masses.  
This parameterization is motivated by galactic feedback, which simulations find evacuates the halo gas below some halo mass threshold (with this threshold affected dependent on how stellar and AGN feedback is implemented).  

\begin{figure}[htpb!]
    \centering
    \includegraphics[width=\linewidth]{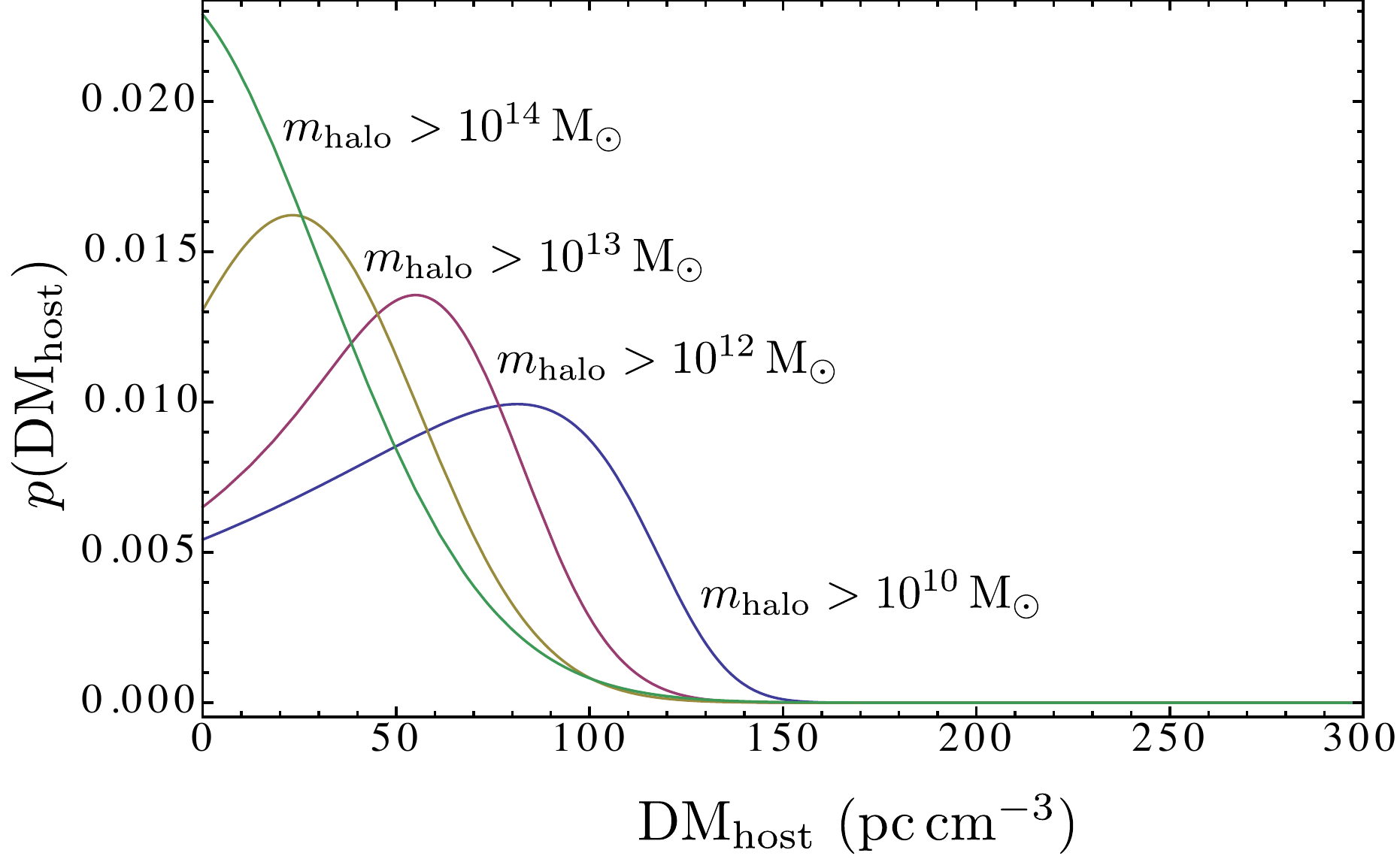}
    \caption{{\bf Posterior probability distributions on the host DM, $p({\rm DM}_{\rm host})$ under a range of assumptions the gaseous profiles around foreground galaxies (see text).}}
    \label{fig:DMhostDist}
\end{figure}

If we posit that ${\rm DM}_{\rm IGM}$ must lie within  the $95\%$ of its allowed range, we obtain a corresponding upper limit on ${\rm DM}_{\rm host}$, after removing the additional contribution from the Milky Way.  The Milky Way contribution lies in the range $55-90$ \pccm, taking into account $40$ \pccm from the Galactic disk, and a further $15-60$ \pccm  due to the (highly) uncertain contribution from the Galactic halo.  Table \ref{tab:dmcontribution} shows the corresponding $95\%$ confidence upper limits on ${\rm DM}_{\rm host}$, for the variety of models considered, assuming a $62$ \pccm dispersion-measure contribution from the Milky Way (i.e., $300$ \pccm\ from the IGM and host combined).  We choose to assume $62$ \pccm for the  the Milky Way (disk and halo) contribution; because it provides conservative (maximal) constraints on the possible host galaxy dispersion contribution. 
The 95\% confidence limit range from 77$(1+z_{\rm host})$ \pccm~to 133$(1+z_{\rm host})$ \pccm. 
We also list the mean expected value of ${\rm DM}_{\rm Host}$ for each of these models.  
The full probability distribution of ${\rm DM}_{\rm host}$, derived from the probability distributions of ${\rm DM}_{\rm IGM}$, is shown in Figure \ref{fig:DMhostDist}.

\subsection{Interpreting the scatter broadening of FRB~180924}

Inhomogeneities in electron density refract and diffract propagating radio waves.  Diffraction is manifested as either scintillation, pulse broadening, and sometimes both.  FRB~180924 shows evidence for spectral modulation like other ASKAP bursts \cite{Macquart18}.  It also shows evidence for pulse broadening.  

The power spectrum of the density fluctuations is modeled to be a power  law between an inner and outer scale of the turbulence:
\begin{equation}
P(q) = C_n^2 q^\beta,   
\end{equation}
where $q$ is the spectral wavenumber and $C_n^2(s)$ represents the amplitude of the turbulence at a distace along the line of sight $s$.  If the turbulence follows a Kolmogorov spectrum, then the power law index is $\beta=-11/3$. 
The scattering measure (SM)  is the integrated value of $C_n^2(s)$ along the line of sight:
\begin{equation}
{\rm SM} = \int ds \, C_n^2. 
\end{equation}
Observable quantities (e.g. the pulse broadening time ($\tau$), scintillation bandwidths, or angular broadening) can be related to the scattering measure assuming a model for the turbulence in the ISM.  

In the Milky Way, for a geometrically thick medium, with density fluctuations following a Kolmogorov spectrum, the scattering measure (SM) can be  calculated from the scatter broadening \cite{NE2001} using
\begin{equation}
{\rm SM}_{\rm MW} = 292\,{\rm kpc\, m^{-20/3}}\,\left(\frac{\tau_{d,s}}{L_{\rm kpc}} \right)^{5/6} \nu_{\rm GHz}^{11/3},
\label{eqn:scattering_eqn}
\end{equation}
where the pulse broadening time ($\tau_{d,s}$) is measured in seconds, the extent of the line of sight ($L_{\rm kpc}$) is measured in kpc, and the observing frequency ($\nu_{\rm GHz}$) is measured in GHz. 
In the Milky Way, other media (such as geometrically thin cases) will result in a relation that changes the coefficient of proportionality by a factor that is close to unity \cite{1999ApJ...517..299L}.

In the case of scattering in the host galaxy, the relationships need to be adjusted to account for the different geometry, and to correct for the redshift of the host. Scattering in the host will be half as effective as would be expected from Equation \ref{eqn:scattering_eqn}.  Additionally, the line of sight length $L$ is measured through the host galaxy (i.e., not the entire path length from the burst source to Earth).  

We then need to convert the scattering measure to the rest frame of the host galaxy:     
\begin{eqnarray}
{\rm SM}_{\rm host} &&=  1560\,{\rm kpc\, m^{-20/3}}\,\left(\frac{2 \tau_{d,s}}{L_{\rm kpc}} \right)^{5/6} \nu_{\rm GHz}^{11/3} (1+z)^{17/3}, \end{eqnarray}
and the average scattering strength along the line of sight is
\begin{eqnarray}
C_{n, {\rm host}}^2 = 1560 \,{\rm m}^{-20/3}\, \tau_{d,s}^{5/6} L_{\rm kpc}^{-11/6} \nu_{\rm GHz}^{11/3} (1+z)^{17/3},
\end{eqnarray}
where $L_{\rm kpc}$ is the length of the line of sight in kpc. 

\begin{figure}[htpb!]
    \centering
    \includegraphics[width=\linewidth]{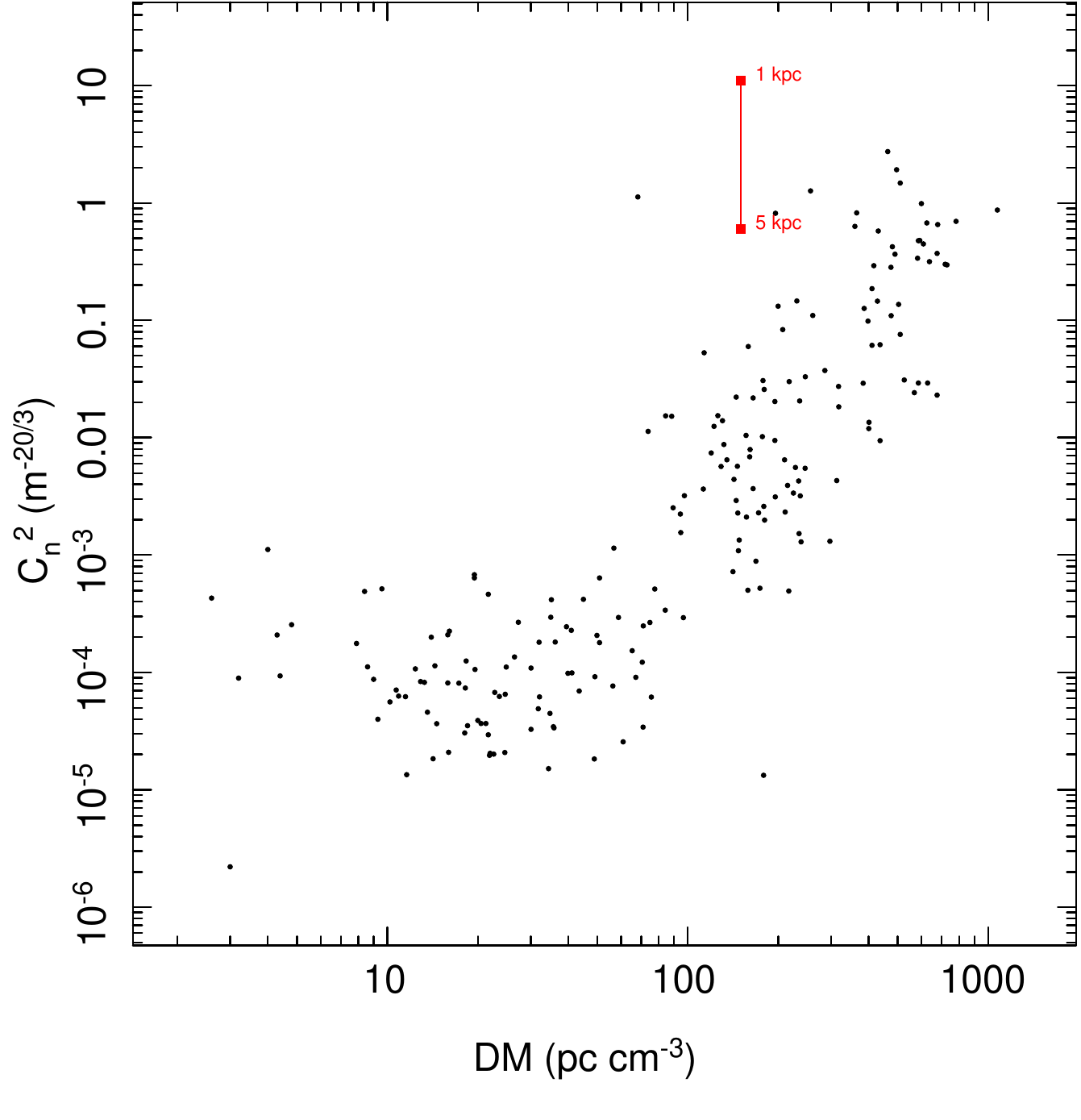}
    \caption{{\bf  Turbulence strength ($C_n^2$) and dispersion measures for Galactic lines of sight} (black points) and FRB~180924 (red). The Galactic lines of sight are from \cite{NE2001II} and \cite{Krishnakumar17}.     }
    \label{fig:cn_pop}
\end{figure}

 Thus for the host of FRB~180924, we find SM=$3 \,{\rm kpc\,m^{-20/3}} \left(L/5\,{\rm kpc} \right)^{-5/6}$, and $C_n^2 = 0.6\,{\rm m^{-20/3}}\left(L/5\,{\rm kpc} \right)^{-5/6}$.
 In Figure \ref{fig:cn_pop}, we show how the scattering strength $C_n^2$ of FRB~180924 compares to measurements along lines of sight to pulsars in the Milky Way, assuming a path length through the host ranging from $1$ to $5$~kpc, and a host dispersion measure of $175$~\pccm\ (in the rest frame) which is the most conservative (largest) 95\% limit from the models considered above.  The plot shows that if the scattering originates in the host, the strength is marginally higher than that measured along Milky Way lines of sight.

\subsection{Acknowledgments}

The Australian Square Kilometre Array Pathfinder, Australia Telescope Compact Array, and Parkes Radio Telescope are part of the Australia Telescope National Facility which is managed by CSIRO. 
Operation of ASKAP is funded by the Australian Government with support from the National Collaborative Research Infrastructure Strategy. ASKAP uses the resources of the Pawsey Supercomputing Centre. Establishment of ASKAP, the Murchison Radio-astronomy Observatory and the Pawsey Supercomputing Centre are initiatives of the Australian Government, with support from the Government of Western Australia and the Science and Industry Endowment Fund. 
Spectra were obtained at the W. M. Keck Observatory, which is operated as a scientific
partnership among Caltech, the University of California, and the National Aeronautics and Space Administration (NASA). The Keck Observatory was made possible by the generous financial support of the W. M. Keck Foundation. 
We acknowledge the Wajarri Yamatji as the traditional owners of the Murchison Radio-astronomy Observatory site. 
The authors recognize and acknowledge the very significant cultural role and reverence that the summit of Mauna Kea has always had within the indigenous Hawaiian community.
We are most fortunate to have the opportunity to conduct observations from this mountain.

The Gemini Observatory is operated by the Association of Universities for Research in Astronomy, Inc., under a cooperative agreement with the NSF on behalf of the Gemini partnership: the National Science Foundation (United States), the National Research Council (Canada), CONICYT (Chile), Ministerio de Ciencia, Tecnolog\'{i}a e Innovaci\'{o}n Productiva (Argentina), and Minist\'{e}rio da Ci\^{e}ncia, Tecnologia e Inova\c{c}\~{a}o (Brazil).
This publication makes use of data products from the Wide-field Infrared Survey Explorer, which is a joint project of the University of California, Los Angeles, and the Jet Propulsion Laboratory/California Institute of Technology, funded by the National Aeronautics and Space Administration.
This project used public archival data from the Dark Energy Survey (DES). Funding for the DES Projects has been provided by the U.S. Department of Energy, the U.S. National Science Foundation, the Ministry of Science and Education of Spain, the Science and Technology Facilities Council of the United Kingdom, the Higher Education Funding Council for England, the National Center for Supercomputing Applications at the University of Illinois at Urbana-Champaign, the Kavli Institute of Cosmological Physics at the University of Chicago, the Center for Cosmology and Astro-Particle Physics at the Ohio State University, the Mitchell Institute for Fundamental Physics and Astronomy at Texas A\&M University, Financiadora de Estudos e Projetos, Funda{\c c}{\~a}o Carlos Chagas Filho de Amparo {\`a} Pesquisa do Estado do Rio de Janeiro, Conselho Nacional de Desenvolvimento Cient{\'i}fico e Tecnol{\'o}gico and the Minist{\'e}rio da Ci{\^e}ncia, Tecnologia e Inova{\c c}{\~a}o, the Deutsche Forschungsgemeinschaft, and the Collaborating Institutions in the Dark Energy Survey.
The Collaborating Institutions are Argonne National Laboratory, the University of California at Santa Cruz, the University of Cambridge, Centro de Investigaciones Energ{\'e}ticas, Medioambientales y Tecnol{\'o}gicas-Madrid, the University of Chicago, University College London, the DES-Brazil Consortium, the University of Edinburgh, the Eidgen{\"o}ssische Technische Hochschule (ETH) Z{\"u}rich,  Fermi National Accelerator Laboratory, the University of Illinois at Urbana-Champaign, the Institut de Ci{\`e}ncies de l'Espai (IEEC/CSIC), the Institut de F{\'i}sica d'Altes Energies, Lawrence Berkeley National Laboratory, the Ludwig-Maximilians Universit{\"a}t M{\"u}nchen and the associated Excellence Cluster Universe, the University of Michigan, the National Optical Astronomy Observatory, the University of Nottingham, The Ohio State University, the OzDES Membership Consortium, the University of Pennsylvania, the University of Portsmouth, SLAC National Accelerator Laboratory, Stanford University, the University of Sussex, and Texas A\&M University.
Based in part on observations at Cerro Tololo Inter-American Observatory, National Optical Astronomy Observatory, which is operated by the Association of Universities for Research in Astronomy (AURA) under a cooperative agreement with the National Science Foundation.

\clearpage

\clearpage

\clearpage

\clearpage

\clearpage

\clearpage

\newpage

\clearpage

 \newpage

\clearpage

\newpage

\FloatBarrier

\section*{Supplementary Tables}

\begin{table}
\caption{{\bf A summary of followup observations of FRB\,180924.} The fourth column shows the number of days since the FRB.} \label{tab:ObsSummary}
\begin{tabular}{|l l l l c|}
\hline
Telescope  & Instrument & UT Date & Days &    Remarks \\
\hline
ASKAP & 1.1-1.4\,GHz  & 2018 Sep 26 & 2 & continuum emission search\\
ATCA & 4.5-8.5\,GHz & 2018 Sep 25 \& Oct 4 & 1 \& 10 & continuum emission search \\
Parkes & 1.4\,GHz & 2018 Oct 2 \& 17  & 8 \& 23 & search for repeat bursts \\
Keck    & KCWI & 2018 Oct 4 & 10 & integral field unit $\lambda \approx 4000-5500$\AA. \\
Gemini-S & GMOS & 2018 Oct 5 & 11 & long-slit spectroscopy $\lambda \approx 4700-9300$\AA \\
VLT & MUSE & 2018 Nov 5 & 41 & $1'\times1'$ integral field unit $\lambda \approx 4750-9300$\AA\\
VLT & FORS2 & 2018 Nov 9 & 45 & $g^{\prime} + I$-band imaging\\ \hline
\end{tabular}

\end{table}

\clearpage

\begin{table}[]
    \centering
        \caption{{\bf Positions and flux densities for the three brightest ($> 10\sigma$) sources in the {\em ASKAP field} data set.} $S_\nu$  is flux density.}
    \label{tab:field_sources}
    \begin{tabular}{lllllll}
    \hline
    Source & R.A.  &  Decl. &   R.A. error & Decl. error  & $S_\nu$ & $S_\nu$  error \\
 & (J2000) &  (J2000) & (mas) & (mas) & mJy  & mJy beam$^{-1}$ .  \\ 
    \hline
PKS~2144$-$418 & 21h44m04.717s  & $-$41d37’51.12” &  90 & 90  & 259 &  3    \\
SUMSS~J214421$-$412640 & 21h44m21.302s  & $-$41d26’40.90” & 280  & 260 & 88 & 3 \\
SUMSS~J214438$-$411835 & 21h44m38.828s  & $-$41d18’33.54” &  943 & 740 &  31 &  3.09 \\
    \hline
    \end{tabular}

\end{table}

\clearpage

\begin{table}[]
    \centering
        \caption{{\bf Polarimetric properties of FRB~180924.}  Stokes parameters, $Q$ and $U$, and the polarization position angle $\Psi$ have been referenced to a frequency of $1.2$~GHz assuming a rotation measure of $14$~rad\,m$^{-2}$.}
    \label{tab:stokes_parameters}
    \begin{tabular}{lllll}
    \hline
    I (Jy)  & Q (Jy) & U (Jy) & V (Jy) & $\Psi$ (rad)  \\ \hline
    11.8(1) &  1.4(1) &-10.5(1) & 1.0 (1) &2.4(1) \\ 
    \hline
    \end{tabular}

\end{table}

\clearpage 

\begin{table}[]
    \centering
        \caption{{\bf Offsets in source positions between Gaia (G) and DES (D)}}
    \label{tab:gaia_des_compare}
    \begin{tabular}{lll}
    \hline
         &  ${\rm RA}_G - {\rm RA}_D$ &  ${\rm DEC}_G - {\rm DEC}_G$ \\ 
         \hline
Without Proper Motion (mas) & 109(20)  & $-$108(17) \\
With Proper Motion (mas) & 102(15) & $-$97(12)\\
\hline
    \end{tabular}

\end{table}

\clearpage

\clearpage

\begin{table}
\centering
\caption{{\bf Host-galaxy photometry} \label{tab:photom}}
\begin{tabular}{ccrc}
\hline 
Survey & Filter & Value & Uncertainty
\\ 
\hline 
DES-DR1 & g & 21.62 & 0.03 \\ 
DES-DR1 & r & 20.54 & 0.02 \\ 
DES-DR1 & i & 20.14 & 0.02 \\ 
DES-DR1 & z & 19.85 & 0.02 \\ 
DES-DR1 & Y & 19.81 & 0.06 \\ 
WISE & W1 & 16.85 & 0.10 \\ 
WISE & W2 & 16.06 & 0.18 \\ 
WISE & W3 & $>11.69$ & $-$ \\ 
WISE & W4 & $>8.50$ & $-$ \\ 
\hline 
\end{tabular} 

\end{table}

\clearpage

\begin{table}
\centering
\caption{{\bf Host Galaxy Nebular Emission}. All magnitudes have been extinction-corrected assuming $A_V=$0.96 mag. \label{tab:lines}}
\begin{minipage}{270mm} 
\begin{tabular}{ccccc}
\hline 
Line & Flux & Flux Uncertainty & Luminosity$^a$ & Lum. Uncertainty \\ 
& ($10^{-17}$ erg/s/cm$^2$) & ($10^{-17}$ erg/s/cm$^2$) 
& ($10^{40}$ erg/s) & ($10^{40}$ erg/s) 
\\ 
\hline 
{[OII] 3726} & 4.0 & 0.2 & 5.6 & 0.3\\ 
{[OII] 3729} & 7.0 & 0.3 & 9.7 & 0.4\\ 
H $\beta$ & 7.3 & 0.2 & 7.4 & 0.2\\ 
{[OIII] 5007} & 7.9 & 0.2 & 7.6 & 0.2\\ 
H $\alpha$ & 28.1 & 0.3 & 20.8 & 0.2\\ 
{[NII] 6583} & 19.5 & 0.3 & 14.4 & 0.2\\ 
{[SII] 6716} & 4.5 & 0.2 & 3.2 & 0.1\\ 
{[SII] 6731} & 3.2 & 0.2 & 2.3 & 0.1\\ 
\hline 
\end{tabular} 
\end{minipage} 
{$^a$}
\label{tab:tblines}
\end{table}

\clearpage

\begin{table}
\centering
\caption{{\bf Results from CIGALE Modeling.}\label{tab:cigale}}

\begin{tabular}{lccc}
\hline 
Parameter & Value & Error 
\\ 
\hline 
Total Mass  ($10^{10} M_\odot$)  & 2.25 & 0.71 \\ 
Old Stellar  Mass ($10^{10} M_\odot$) & 2.24 & 0.71  \\ 
AGN Fraction & 0.2 & 0.2  \\ 
$u-r$ (rest frame) & 1.7 & 0.2 \\ 
$M_r$ (rest frame) & -20.77 & 0.05 \\ 
SFR $M_\odot \, \rm yr^{-1}$  & 2.9 & 3.9 \\ 
$E(B-V)$  & 0.27 & 0.11  \\ 
$t_{\rm age}$ (Gyr) & 5.55 & 3.17 \\ 
\hline 
\end{tabular} 
\end{table}

\clearpage
\begin{table}[]
    \centering
        \caption{{\bf Location of stars in the nearby Universe} derived from the galaxy stellar mass function \cite{baldry2012}.  }
    \label{tab:stars}
    \begin{tabular}{lr}
    \hline
Galaxy stellar mass (M$_\odot$) & Fraction of local stars \\
\hline
    10$^{7}$ to 10$^{8}$  & 1\% \\
10$^{8}$ to 10$^{9}$ & 4\% \\
10$^{9}$ to 10 $^{10}$ & 16\% \\
10$^{10}$ to 10$^{11}$ & 68\% \\
10$^{11}$ to 10$^{12}$  & 11\%\\
         \hline 

\end{tabular}

\end{table}

\clearpage

\begin{table}[]
    \centering
        \caption{{\bf Host DM likelihoods for $z=0.32$ and an upper limit of 300 \pccm.}}
    \label{tab:dmcontribution}
    \begin{tabular}{lllll}
    \hline
      & Halo Radius & Mean& Standard deviation & $95\%$ DM \\
      & ( $R_{\rm vir}$) &  (\pccm) & (\pccm) & (\pccm) \\ 
      \hline
      Top-hat model & 0.5 & 81 & 39 &  133\\
      & 0.75&  62 & 36 &  122 \\
      & 1&  51 & 32 & 108 \\
      & 2 &  32 & 24 &  79\\
      \hline
      NFW model & 1 & 66 & 34& 119 \\
      & 2 & 43 & 29 &  95 \\ 
      \hline \hline
      &  Minimum halo mass &  Mean & Standard Deviation &  $95\%$ DM \\
      & ($M_{\cdot}$) & (\pccm) &  (\pccm)  &  (\pccm) \\ \hline
      NFW Model  
       &  $10^{10}$ & 66 & 34 & 119\\
     extended to   &  $10^{12}$ & 50 & 27 & 94 \\
     $1.0\times R_{\rm virial}$  &  $10^{13}$ & 36 & 23 & 79 \\
       &  $10^{14}$ & 30 &  24 &  77 \\
       \hline
      
    \end{tabular}

\end{table}

\clearpage

\end{document}